\documentclass[11pt,reqno]{amsart}

\usepackage{amsmath}
\usepackage{amsfonts}
\usepackage{amssymb}
\usepackage{amsthm}
\usepackage{bbm}
\usepackage{newlfont}
\usepackage{xcolor}

\usepackage{enumerate}

\usepackage{graphicx}

\theoremstyle{definition}
\newtheorem{defn}{Definition}[section]
\theoremstyle{plain}
\newtheorem{thm}[defn]{Theorem}

\newtheorem{tvr}[defn]{Proposition}

\theoremstyle{remark}
\newtheorem{remark}{Remark}

\newcommand{\R}{\mathbb{R}}
\newcommand{\Z}{\mathbb{Z}}
\newcommand{\C}{\mathbb{C}}
\newcommand{\N}{\mathbb{N}}
\newcommand{\T}{\mathbb{T}}
\newcommand{\G}{\mathbb{G}}
\newcommand{\set}[2]{\left\{#1  \mid #2 \right\}}
\newcommand{\setcomb}[2]{
 \begin{smallmatrix}
#1 \\ #2
\end{smallmatrix}  }

\allowdisplaybreaks
\usepackage{backref}

\begin{document}

\title[]{Gaussian cubature arising from hybrid characters of simple Lie groups}

\author{R.~V.~ Moody$^1{}^,{}^4$}
\author{L.~Motlochov\'{a}$^3$}
\author{J.~Patera$^2{}^,{}^4$}

\begin{abstract}\

Lie groups with two different root lengths allow two `mixed sign' homomorphisms
on their corresponding Weyl groups, which in turn give rise to two families of hybrid Weyl group orbit functions and characters. In this paper we extend the ideas leading to the Gaussian cubature formulas for families of polynomials arising from the characters of irreducible representations of any simple Lie group, to new cubature formulas based on the corresponding hybrid characters. These formulas are new forms of Gaussian cubature in the short root length case and new forms of Radau cubature in the long root case. The nodes for the cubature arise quite naturally from the (computationally 
efficient) elements of finite order of the Lie group.
\end{abstract}\

\date{\today}

\maketitle

\noindent
$^1$ Department of Mathematics and Statistics, University of Victoria, Victoria, BC., V8W\,3R4 Canada
\\
$^2$ Centre de recherches math\'ematiques, Universit\'e de Montr\'eal, C.~P.~6128 -- Centre ville,
Montr\'eal, H3C\,3J7, Qu\'ebec, Canada
\\
$^3$ D\'epartement de math\'ematiques et de statistiques, Universit\'e de Montr\'eal, C.~P.~6128 -- Centre ville,
Montr\'eal, H3C\,3J7, Qu\'ebec, Canada
\\
$^4$ MIND Research Institute, 111 Academy Drive, Irvine, California 92617

\noindent $\phantom{^\S}$~E-mail: rmoody@mac.com, motlochova@dms.umontreal.ca, patera@crm.umontreal.ca

\bigskip


\smallskip


\section{Introduction}
It has long been known that the Chebyshev polynomials of the second kind are related to the 
representation theory of $SU(2)$, and of course to efficient methods of numerical quadrature. In \cite{LX} it was shown that there is a considerable generalization of this theory based
on the series of lattices of type $A_{n}$ (so that the original theory applied to the lattice of type $A_{1}$ and the representations of
$SU(2)$). This generalization depended deeply on the Weyl groups of these
lattices, but not particularly on the Lie groups associated with them. The resulting formulas, now
for functions of $n$ variables, went under the name of cubature formulas.

In \cite{MPCubature} the idea that there is a genuine Lie theoretical connection here was extended to create
a theory that works for every simple compact Lie group $\G$. The theory is again based on the root lattices
but now also incorporates the representation theory of these groups in a deeper way,
and more importantly uses the elements of finite order in the corresponding Lie group to define the 
nodes at which the cubature formulae are evaluated. The representations and the elements of finite order
are in a sort of duality, and this duality plays a vital role in what happens. With a slight Lie-theoretical twist in the definition of the degrees of multi-variable polynomials, the crucial polynomials, their nodes
and the cubature formulas appear completely naturally out of the theory and in fact are optimal (called Gaussian) in their efficiency.

The Weyl group $W$, which appears as a group of {\em reflections} in this theory, is of primary importance,
notably its sign homomorphism $W \longrightarrow \{\pm 1\}$
which takes the sign $-1$ for each of the reflections in the roots. It has long been 
known in the theory of orthogonal polynomials based on these reflection groups that in the cases where the simple
Lie group has roots of two different lengths (namely for types $B_{n},C_{n},F_{4},G_{2}$) there
are, in addition, two hybrid sign functions which distinguish between reflections in long roots and reflections in short
roots; that is, the sign function takes the value $-1$ for each reflection in a long root (respectively short root) 
and takes the value $+1$ on the reflections in the short (respectively long) roots. 

In this paper we extend the ideas of Chebyshev polynomials, nodes, and cubature formulas to these hybrid situations. In principle the path should be straightforward, particularly since orthogonal polynomials and $q$-series based on this type of hybrid symmetry have been well studied, e.g.
\cite{hec}. 
However, our theory depends on both the representations and the
elements of finite order of the Lie group, and this somewhat intricate process requires making a number of correct decisions in how to define things to fit the new setting. In the end things work out as smoothly and as naturally as in \cite{MPCubature}, although
for the long root case the cubature is slightly less efficient than in the Gaussian cubature of the standard and short root cases, being instead what is called Radau cubature.

The orientation of \cite{MPCubature} was towards the approximation theory community
since Gaussian formulas are rather rare and the Lie theoretical connections offer new and
unexpected 
techniques for constructing them. In this paper, in addition to presenting the new results based
on hybrid Weyl symmetry and simplifying the overall presentation of the ideas, the emphasis is more the other way around, aiming to introduce the Lie theoretical community to some new applications of simple Lie groups to approximation theory and cubature. It seems to us that there is more to be explored here, particularly the duality between
elements of finite order and character theory.

\section{Overview}
We begin with a summary of the results of \cite{MPCubature} and then introduce the ideas which lead to the new cubature formulas arising from the two new families of orbit functions. 

Start with the polynomial ring $\C[X_{1}, \dots, X_{n}]$. This is given the structure of a graded ring by assigning a degree $d_{j} \in \Z^{>0}$ 
(called the $m$-degree, for reasons to be explained later) to each of the variables $X_{j}$. The degree of a monomial 
$X_{1}^{k_{1}} \cdots X_{n}^{k_{n}}$ is thus $k_{1}d_{1} + \dots
+ k_{n}d_{n}$. Unlike the usual gradation, $d_{j}$ need
not be equal to $1$.
The value of $n$ will ultimately be the rank of a compact
simple Lie group $\G$ (or its complex simple Lie algebra
$\mathfrak g$) and the degree structure will be given by the coefficients
of its highest co-root. 

The main result can be stated as a quadrature formula, called
in this subject a {\bf cubature formula} because it is not restricted
to one dimension. Fix any non-negative
integer $M$. Then for all $f \in \C[X_{1},\dots X_{n}]$ of 
$m$-degree not exceeding $2M+1$, 
\begin{equation}\label{mainFormulaSimplified}
(2 \pi)^{-n}\int_{\Omega} f(X) K^{1/2}(X) dX
= C \sum_{X\in \mathcal F_{M+h}} f(X) K(X) \,.
\end{equation}
The main point is that integration is replaced
by finite summing, and the elements of $\mathcal F_{M+h}$ over
which the summation takes place are very easy to compute. Here $X = (X_{1}, \dots, X_{n}) \in \C^{n}$ and $\mathcal F_{M+h}$ is a finite subset of $\C^{n}$, $C$ is a constant, $K$ is a special polynomial in $\C[X_{1}, \dots, X_{n}]$ which is positive valued on 
$\Omega \subset \C^{n}$. All of these objects depend on the choice of 
$\G$. In the hybrid situation that we shall develop here, the variables $X^{s} = (X^{s}_{1}, \dots, X^{s}_{n})$ and similarly $X^{l} = (X^{l}_{1}, \dots, X^{l}_{n})$ are real valued. 

The elements of $\mathcal F_{M+h}$ actually arise from elements of $\G$ finite order,
but in this context they are called the {\bf nodes}, and they have
a number of special properties.  Their number
is exactly the dimension of the space of polynomials of
$m$-degree at most $M$. Furthermore, an important part of the construction of this
result is the introduction of special polynomials (related to characters
and other $\G$-invariant functions on $\G$)
$X_{\lambda}= X_{(\lambda_{1},\dots, \lambda_{n})}$ of $m$-degree
$|\lambda|_{m}:=\lambda_{1}d_{1} + \cdots + \lambda_{n}d_{n}$, which form an orthogonal
basis of $\C[X_{1}, \dots, X_{n}]$ with respect to the inner product
\begin{equation}\label{innerProductSimplified}
\langle f, \overline g\rangle_{K} := (2 \pi)^{-n}\int_{\Omega} f \overline g K^{1/2} \,,
\end{equation}
which in view of \eqref{mainFormulaSimplified} is 
$\sum_{X\in \mathcal F_{M+h}} f(X) \overline {g(X)} K(X)$ if the $m$-degrees
of $f,g$ do not exceed $M$. Now, the minimum number of nodes that could
achieve such an orthogonal decomposition of these functions is the
dimension of the space of polynomials of $m$-degree at most $M$, and that
is exactly the number of elements in $\mathcal F_{M+h}$. This optimal situation is called {\bf Gaussian cubature} \cite{LX}.

The nodes are actually zeros of certain of these polynomials of degree $M+1$.
The region $\Omega$ is the image of the interior of the fundamental region (or some modified
version of it in the hybrid cases) under a certain polynomial map. In particular it
is an open set with compact closure and boundary of measure $0$.

If we move to the Hilbert space $L^{2}_{K}(\Omega)$ of square
integrable functions on $\Omega$ with respect to the inner product
$\langle \cdot, \cdot \rangle_{K}$ then every function $f \in L^{2}_{K}(\Omega)$ has a Fourier expansion
\begin{equation}\label{FourierSimplified}
f = \sum_{\lambda} \langle f, X_{\lambda}\rangle_{K} X_{\lambda}\,,
\end{equation}
equality here being in the usual $L^{2}$ sense.
If the sum is truncated to 
$\sum_{|\lambda|_{m}\le M}\langle f, X_{\lambda}\rangle X_{\lambda}$
then this is the best approximation to $f$ in the $L^{2}_{K}$-norm 
using only polynomials of $m$-degree at most $M$.  

In essence what we have been describing arises from a duality that exists between the characters
of the representations of $\G$ and the conjugacy classes of elements of finite
order of $\G$. Let $\T$ be a maximal torus of $\G$. Since all the maximal tori are 
conjugate and every conjugacy class of $\G$ meets every one of them, 
every character of $\G$ is defined entirely
by its restriction to $\T$ and every conjugacy class of elements of finite order
has elements in $\T$. The relationship between $\G$ and its Lie algebra
restricts to the relationship between $\T$ and its Lie algebra: 
\begin{equation}\label{exp1}\exp{2\pi i (\cdot)}: \mathfrak{t}\rightarrow\mathbb{T}\,.\end{equation}

Here it is more convenient to let $i\mathfrak t$ be the Lie algebra of $\T$
because the Killing form is then positive definite on $\mathfrak t \simeq \R^{n}$,
where $n$ is the rank of $\G$.
The kernel of this exponential mapping is the co-root lattice $Q^{\vee}$ of $\G$,
so $\T \simeq \R^{n}/Q^{\vee}$. The $\Z$-dual of $Q^{\vee}$ in $\mathfrak t^{*}$
is the weight lattice $P$.

The normalizer $N$ of $\T$ in $\G$ is always
larger than $\T$ itself, and the Weyl group $W:=N/\T$ is the group
that represents this excess. $W$ acts  on $\T$ via conjugation
 and then as linear transformations on $\mathfrak t$. The affine Weyl group is then 
 the  semi-direct product of $W_{\mathrm{aff}}=W\ltimes Q^\vee$, which acts
 on $\mathfrak t$ with $Q^{\vee}$ acting as translations.
 
 Let $F$ be a standard simplicial fundamental region for $W_{\mathrm{aff}}$
in $\mathfrak t$, so that $W_{\mathrm{aff}}$ is generated by the reflections
in the faces of $F$ and $W$ is generated by the reflections in the
faces of $F$ that pass through the origin, see \cite{B}. The virtue of $F$
is that it perfectly parametrizes the conjugacy classes of $\G$: for each
such class there is a unique element of $x \in F$ for which $\exp(2\pi i x)$ lies in that class.
 
 The characters on $\G$ restrict faithfully to $W$-invariant functions on $\T$,
 and the ring of all $W$-invariant functions on $\T$ is a polynomial
 ring in $n$-variables generated by the characters of a set of so-called fundamental
 representations. This is the ring $\C[X_{1}, \dots X_{n}]$ and the
 $X_{j}$ can be viewed either abstractly as variables or as actual characters
 corresponding to a system of fundamental weights. One particularly important
$W$-invariant function on $\T$ is $K:= |S_{\rho}|^{2}$ where 
$S_{\rho}$ is the basic skew-symmetric function that appears as the denominator
of Weyl's character formula. This is the $K$ of the cubature formula.
  
Via the exponential mapping the characters can be viewed as $W_{\mathrm{aff}}$-invariant functions on $\mathfrak t$. In this way we have the important mapping
\begin{equation}\label{polyMap}
\Xi: \mathfrak t \longrightarrow \C^{n} \qquad  x \mapsto (X_{1}(\exp(2\pi i x), \dots,
X_{n}(\exp(2\pi i x))
\end{equation}

The region $\Omega$ is the image of the interior $F^{\circ}$ of $F$ under $\Xi$. 

\begin{remark} \label{xXremark} There are several points of possible confusion
regarding the many functions that appear in the paper. First of all there are
many functions, like $S_{\rho}$, which 
have interpretations as functions both on $\T$ and on $\mathfrak t$.
This is not particularly troublesome since
$\T \simeq \R^{n}/Q^{\vee}$ and all these functions are clearly periodic on 
$\mathfrak t$ with respect to $Q^{\vee}$. Thus interpreting 
$S_{\rho}\overline{S_{\rho}}$ as a function on $\mathfrak t$ or $\T$ is rather obvious.

The second is the transition from exponential sums to 
new coordinates in $\C^{n}$ using characters (or hybrid characters) as new variables. 
This is the way in which the Lie theory
translates over into a theory about polynomials where the cubature formulas
are relevant. Rather than introduce new function names when we transition variables,
we use different notation for the variables. Thus for functions on $\mathfrak t$ or $\T$ the
generic variable name is $x= (x_{1}, \dots, x_{n}) \in \R^{n}$, whereas for the new
polynomial variables the generic variable name is $X=(X_{1},\dots, X_{n})\in \C^{n}$. 
When we deal with short and long root scenarios, as we mostly
do in what follows, we use $X^{s}=(X^{s}_{1},\dots, X^{s}_{n})$ in the short root case, and similarly
for the long case. 
\end{remark}

There remains to briefly introduce the elements of finite order of $\T$. Each conjugacy
class of an element of finite order has a unique representative in $F$. The set 
$\mathcal F_{M+h} \subset \Omega$ is the image under $\Xi$ of the
set of elements in $F$ that
have adjoint order $M+h$. Here $h$ is the Coxeter number of $\G$ and
by adjoint order we mean that the order of the element is $M+h$ in the
adjoint representation of $\G$ on itself (i.e. by conjugation). The full
order of an element is a finite multiple of the adjoint order.

This finishes our brief tour of the constituents of the basic cubature formula.

The Weyl group is a subgroup of the orthogonal group of $\mathfrak t$ with
respect to its canonical Euclidean structure arising from the Killing form, and
in particular there is the sign homomorphism
\[ \sigma : W \longrightarrow \{\pm1\}  \qquad w\mapsto \sigma(w) = \det(w) \, ,\]
with $\sigma(r) =-1 $ for all reflections.
The fact that $W$ is generated by the reflections in the roots of the Lie algebra
plays an essential role in elucidating the structure of simple Lie groups and their representations. 
Throughout, $W$-skew invariant functions and polynomials play a key role, Weyl's character 
formula being a typical example which expresses the characters ($W$-invariant exponential
sums) as ratios of $W$-skew invariant exponential sums.
In the case when the roots of the Lie algebra have two distinct lengths (called
the short and long roots), there are two alternative
hybrid sign homomorphisms:  $\sigma^{s}$ which is defined by taking the value
$-1$ on the reflections in short roots and the value $+1$ on the reflections in long roots,
and $\sigma^{l}$ which does it the other way around. This gives rise to new 
hybrid invariants, skew invariant with respect to short reflections while being invariant with respect to long, or vice-versa. This leads to two new versions of each cubature formula, see \eqref{slcubature} which say very much the same thing except that 
$\Omega, K, \mathcal F_{M+h}, C$ and a new function $\kappa$, all appear
in short and long forms according to which 
hybrid symmetry is used.  The effect is somewhat subtle: $\Omega$ is only altered along its boundary,
the set $\mathcal F_{M+h}$ changes only by certain elements of finite order along the 
boundary of the fundamental region $F$, and the polynomial ring is still the space of $W$-invariant functions. However 
the interpretations of the variables $X_{j}$ in terms of characters and the function $K$ are 
significantly altered.

\section{Basics}\label{Basics}
 
 We establish the notation that we are using and recall some basic facts about simple Lie algebras. For more details, see for example \cite{mpsk}. 
 
\subsection{Simple Lie algebras}\label{simpleLie}\

Let $\mathfrak{g}$ be a simple complex Lie algebra of rank $n$ with corresponding simple and simply connected compact Lie group  $\G$.  Let $\mathbb{T}$ be a maximal torus of $\G$ and 
let $i\mathfrak{t}$ be its Lie algebra, so that we have the exponential map \eqref{exp1}.
Let $(\cdot\mid\cdot)$ on the dual space $\mathfrak{t}^*$
 of $\mathfrak{t}$ be defined from the Killing form by duality. The natural pairing of $\mathfrak{t}^*$ and $\mathfrak{t}$ is denoted by $\langle\,\cdot\,,\cdot\,\rangle$.
 
Let $\Delta$ denote the set of roots of $\mathfrak{g}$ and let $\Pi:=\{\alpha_1,\dots,\alpha_n\} \subset \mathfrak{t}^*$
be a set of simple roots, hence also a basis of $\mathfrak{t}^*\simeq\R^n$.
We denote by $C$ the corresponding Cartan matrix with entries
\begin{equation*}
C_{ij}=\frac{2(\alpha_i\mid\alpha_j)}{(\alpha_j\mid\alpha_j)}\,.
\end{equation*} 
Its determinant, denoted by $c_\mathfrak{g}$, is the order of the centre of $\G$ and
is also the index of the root (co-root) lattice in side the weight (co-weight) lattice, see below.

We introduce the usual partial ordering on $\mathfrak{t}^*$: 
$\mu\preceq\lambda$ if and only if $\lambda-\mu$ is a sum of simple roots or $\lambda=\mu$.   The highest root in $\Delta$ with respect to this ordering is denoted $\xi$. Its coordinates in the $\alpha$-basis are
called the marks:
\begin{equation}\label{marks}
\xi=m_1\alpha_1+\dots+m_n\alpha_n\,.
\end{equation}

Let
$Q, P\subset \mathfrak{t}^*$ be the root lattice and weight lattice respectively.
Then
\begin{equation*}
P=\set{\lambda\in\mathfrak{t}^*}{\langle\lambda,\alpha_j^\vee \rangle \in \Z\text{ for }\forall\alpha_j^\vee, j= 1, \dots, n}\,,
\end{equation*}
where $\Pi^{\vee}:= \{\alpha_{1}^{\vee}, \dots, \alpha_{n}^{\vee}\}$ is the system of simple co-roots (which forms a basis in $\mathfrak{t}$) defined by
\begin{equation*}
\langle\alpha_i,\alpha_j^\vee\rangle= C_{ij}\qquad \text{for }i,j=1,\dots,n\,.
\end{equation*} 
To these simple co-roots corresponds the system of co-roots $\Delta^\vee$, which is in fact the system of roots for the simple Lie algebra with Cartan matrix $C^T$ (although
this algebra never makes any real appearance in what follows). We have the highest co-root in $\eta \in \Delta^\vee$ and giving the co-marks $m_j^\vee$:
\begin{equation*}\label{comarks}
\eta=m_1^\vee\alpha_1^\vee+\dots+m_n^\vee\alpha_n^\vee\,.
\end{equation*}
It is these co-marks that define the degree function on $\C[X_{1}, \dots, X_{n}]$ later.

The lattice $P$ has as a basis the set of fundamental weights $\omega_i$ which is dual to the co-root basis in the sense that
\begin{equation*}
\langle\omega_i,\alpha_j^\vee\rangle=\delta_{ij}\qquad \text{for }i,j=1,\dots,n \,.
\end{equation*}
This is so called $\omega-$basis of $\mathfrak{t}^*$ that we will use.

We also have two lattices in $\mathfrak{t}$ denoted $Q^\vee$ and $P^\vee$. The co-root lattice $Q^\vee$ is  kernel of the exponential map \eqref{exp1} with $\Z$-basis consisting of the $\alpha_i^\vee$.  
The co-weight  lattice $P^\vee$ is the $\Z$-dual of $Q$ in $\mathfrak{t}$ and has as a basis the set of fundamental co-weights $\omega_j^\vee$ defined by
\begin{equation*}
\langle \alpha_i,\omega_j^\vee\rangle=\delta_{ij}\qquad \text{for }i,j=1,\dots,n\,.
\end{equation*}

The relationships between the lattices and between the various root and weight bases and their co-equivalents described below are summarized in:

\begin{equation*}
\begin{matrix}
\{\alpha_1, \dots, \alpha_n\}
  &\subset 
  &Q&&Q^\vee
  &\supset
  &\{\alpha^\vee_1, \dots, \alpha^\vee_n\}\\
 &&\cap&\times&\cap\\
\{\omega_1, \dots,\omega_n\}
  &\subset 
  &P&& P^\vee&\supset&\{\omega^\vee_1, \dots, \omega^\vee_n\}\\
&&\cap&&\cap\\
&&\mathfrak{t}^*&&\mathfrak{t}
\end{matrix}
\end{equation*}
Here the times symbol is meant to indicate that $Q$ and $P^\vee$, as well as $P$ and $ Q^\vee$, are in $\Z$-duality with each other.

Finally we have the cone $P^+ \subset P$ of dominant weights: 
\begin{equation*}
P^+=\Z^{\geq0}\omega_1+\dots+\Z^{\geq0}\omega_n\,.
\end{equation*}

\subsection{Affine Weyl group and its dual}\

The Weyl group acting on $\mathfrak{t}$ is generated by simple reflections $r_1,\dots, r_n$ in the hyperplanes
\begin{equation*}
H_i:=\set{x\in\mathfrak{t}}{\langle\alpha_i,x\rangle=0}\,,\qquad i=1,\dots,n
\end{equation*}
by
\begin{equation*}
r_i(x):=x-\langle\alpha_i,x\rangle\alpha_i^\vee\,.
\end{equation*}
By duality, we have the action of $W$ on $\mathfrak{t}^*$ where the simple reflections on co-root side are given by
\begin{equation*}
r_i(\lambda):=\lambda-\langle\lambda,\alpha_i^\vee\rangle\alpha_i\,.
\end{equation*}

The affine Weyl group is the semi-direct product of $W$ and the translation group $Q^\vee$:  $W_{\mathrm{aff}}=W\ltimes Q^\vee$. Equivalently, $W_{\mathrm{aff}}$ can be defined as the group generated by the simple reflections $r_i$ and the affine reflection $r_0$ given by
\begin{equation*}
r_0(x)=r_\xi(x)+\xi^\vee\,,\qquad r_\xi(x)=x-\langle\xi,x \rangle\xi^\vee
\end{equation*}
where $\xi$ is the highest root of $\Delta$. 
 
The standard simplex $F$ in $\R^n$ defined by
\begin{equation*}
F=\{x\mid \langle\alpha_j,x\rangle\geq0
\quad\text{ for all }\ 
j=1,\dots,n,\quad \langle\xi,x\rangle\leq1\}\,,
\end{equation*}
serves as a fundamental domain for the affine Weyl group.
Its vertices are
\begin{equation}\label{funddom}
F=\left\{0,\tfrac1{m_1}\omega_1^\vee,  
         \dots, 
         \tfrac1{m_n}\omega_n^\vee\right\}\,,
\end{equation}
where $m_i$, $i=1,2,\dots,n$, are the marks \eqref{marks}. Note that $r_0$ is the reflection in the hyperplane $H_0$ 
\begin{equation}\label{stena0}
H_0:=\set{x\in\mathfrak{t}}{\langle\xi,x\rangle=1}\,.
\end{equation} 

\subsection{Long and short roots}\ 

In dealing with the hybrid cases, we are only interested in the simple Lie algebras 
with two different lengths of roots:
\begin{equation*}
B_{n}\ (n\geq3)\,,\qquad C_n\ (n\geq2)\,,\qquad F_4\,,\qquad G_2\,.
\end{equation*}
The root system $\Delta$ of such algebras consists of short roots $\Delta^s$ and long roots $\Delta^l$, so \begin{gather}
\Delta=\Delta^l\cup\Delta^s\,.
\end{gather}
Similarly, we decompose the set of simple roots $\Pi$ as $\Pi=\Pi^l\cup\Pi^s$ where $\Pi^l:=\Pi\cap\Delta^l$ and $\Pi^s:=\Pi\cap\Delta^s$.
Our indexing of the simple roots is such that
\begin{equation*}
\begin{alignedat}{2}
\Delta^l(B_n) &\ni\alpha_1,\dots,\alpha_{n-1} &\qquad
\Delta^l(C_n) &\ni\alpha_n,\\
\Delta^l(F_4) &\ni\alpha_1,\alpha_2&\qquad \qquad
\Delta^l(G_2) &\ni\alpha_1\,.
\end{alignedat}
\end{equation*}

Since $\Delta^s$ and $\Delta^l$ are stabilized by $W$ and span 
$\mathfrak{t}^*$, they both form root systems in \ $\mathfrak{t}^*$. Although we do not use the facts here, it is known that $\Delta^l$ is the root system of a semisimple subalgebra of the simple Lie algebra $\mathfrak g$ belonging to $\Delta$ and $\Delta^s$ is the root system of a subjoined semisimple Lie algebra \cite{PSS,MPi}, which is usually not a subalgebra of $\mathfrak g$.
\begin{equation}\label{subroots}
\Delta^s\text{ is of type }
\begin{cases}
   nA_1\ \text{in}\ B_n\\
   D_n\  \text{in}\ C_n\\
   D_4\  \text{in}\ F_4\\
   A_2\  \text{in}\ G_2\\  
\end{cases}\,,\qquad\qquad
\Delta^l\text{ is of type }
\begin{cases}
   D_n\  \text{in}\ B_n\\
   nA_1\ \text{in}\ C_n\\
   D_4\  \text{in}\ F_4\\
   A_2\  \text{in}\ G_2\\  
\end{cases}\,,
\end{equation}
where $nA_1$ denotes the semisimple Lie algebra, $nA_1=A_1\times\cdots\times A_1$, ($n$
factors). In \eqref{subroots} we use the isomorphisms $D_2\simeq A_1\times A_1$ and $D_3\simeq A_3$.

Define the set of positive short and positive long roots by
$\Delta^s_+:=\Delta^s\cap\Delta_+,\,\Delta^l_+:=\Delta^l\cap\Delta_+$ respectively.

\begin{tvr}\ 
$\Delta^t_+$ \ is a system of positive roots for $\Delta^t$ where $t\in\{s,l\}$.
\end{tvr}

\begin{proof}
All systems of positive roots in any root system $\Sigma$ arise as
$\Sigma_+=\set{\alpha \in \Sigma}{(\nu\mid\alpha)>0}$
for some $\nu$ in the span of $\Sigma$ \cite{Serre}. Now with $\rho$ being half the sum
of the positive roots of $\Delta$, we have $\Delta_+=\set{\alpha\in\Delta}{(\rho\mid\alpha)>0}$. Then $\Delta^t_+=\set{\alpha\in\Delta^t}{\alpha\in\Delta_+}=
\set{\alpha\in\Delta^t}{(\rho\mid\alpha)>0}$. So $\Delta^t_+$ is a positive root system. 
\end{proof}

The highest long root $\gamma^l$ of $\Delta^l$ coincides with the highest root $\xi$ of $\Delta$. So, the coefficients of $\gamma^l$ written in $\alpha-$basis are the marks $m_i$, $\gamma^l=m_1\alpha_1+\dots+m_n\alpha_n$, see Table \ref{rootnumbers}. The highest short root of $\Delta^s$ denoted $\gamma^s$ is given by its coefficients $m_i^s$ in $\alpha-$basis, $\gamma^s=m_1^s\alpha_1+\cdots+m_n^s\alpha_n$, see Table \ref{rootnumbers}.

The dual root system $\Delta^\vee$ decomposes also as disjoint union of short co-roots $\Delta^{\vee s}$ and long co-roots $\Delta^{\vee l}$. The dual of $\gamma^l$ is the highest short co-root $\gamma^{l\vee}=m_1^{l\vee }\alpha_1^\vee+\cdots+m_n^{l\vee }\alpha_n^\vee$.
Note: we label the highest short root with `l' to express the duality with the highest long root. Similarly, the dual of $\gamma^s$ is the highest long co-root $\gamma^{s\vee}=
m_1^\vee\alpha_1^\vee+\dots+m_n^\vee\alpha_n^\vee$. The values of $m_i^\vee$ and $m_i^{l\vee}$ are written in Table \ref{rootnumbers}.

\begin{table}[ht] 
\footnotesize
\addtolength{\tabcolsep}{-3pt}

\begin{center}
\begin{tabular}{|c||c|c|c|c|c|c|c|c|}
\hline
$\Delta$ &$ \ m_1,\dots,m_n\ $  
        &$ \ m_1^s,\dots,m_n^s\  $   
        &$ \ m_1^{l\vee },\dots,m_n^{l\vee }\ $
        &$ \ m_1^\vee,\dots,m_n^\vee\ $
        &$ \rho^l $
        &$ \rho^s $
        &$ h^l $
        &$ h^s $
\\\hline\hline 
$B_n$   &$1,2,\dots,2$ 
             &$1, \dots, 1$
             &$1, 2,\dots,2, 1$
             &$2,2, \dots, 2, 1$      
             &$1, \dots, 1,0$
             &$0, \dots,0,1$
             &$2n-2$
            &$2$
\\\hline
$C_n$  &$2,\dots,2,1$ 
             &$1,2,\dots,2, 1$
             &$1,\dots, 1$
             &$1,2, \dots, 2$       
             &$0,\dots, 0,1$
             &$1, \dots,1,0$
             &$2$
            &$2n-2$
\\\hline
$F_4$  &$2,3,4,2$ 
             &$1,2,3,2$             
             &$2,3,2,1$
             &$2,4,3,2$
             &$1,1,0,0$
             &$0,0,1,1$
             &$6$
            &$6$
 \\\hline
$G_2$  &$2,3$ 
             &$1,2$             
             &$2,1$
             &$3,2$
             &$1,0$
             &$0,1$
             &$3$
            &$3$

 \\\hline\hline 
 
\end{tabular}
\bigskip
\caption{The numbers $m_i$ and $m_i^s$  are the coefficients of the highest long root $\gamma^l$ and highest short root $\gamma^s$, written in the
standard basis of simple roots. Similarly, ${m_i^l}^\vee$ and ${m_i}^\vee$ 
are the coefficients of the duals of the $\gamma^l$ and $\gamma^s$, written in the basis of simple co-roots. As for $\rho^l$ and $\rho^s$, these columns are the coefficients of the half-sums of the positive long and
short roots, written in the basis of fundamental weights. Finally, $h^s$ and $h^l$ denote the numbers \eqref{hshl}. }
\label{rootnumbers}
\end{center}
\end{table}

\bigskip

A function
\begin{equation}\label{mult}
k:\alpha\in\Delta\rightarrow k_\alpha\in\R
\end{equation}  
for which $k_\alpha=k_{w(\alpha)}$ for $w\in W$ is called
a multiplicity function \cite{hec}. The trivial example is $k_\alpha=1$ for all $\alpha\in\Delta$ which we denote simply by $k_0$. Relevant for us 
are
\begin{equation}\label{kfunction}
\begin{aligned}
&k^l:\quad k^l_\alpha:=1 \quad \text{for } \alpha\in\Delta^l \quad\text{and}\quad  k^l_\alpha:=0 \quad \text{for } \alpha\in\Delta^s,  \mbox{and}\\
&k^s:\quad k^s_\alpha:=0 \quad \text{for } \alpha\in\Delta^l \quad\text{and}\quad  k^s_\alpha:=1 \quad \text{for } \alpha\in\Delta^s\,.
\end{aligned}
\end{equation}

 Defining
\begin{equation}\label{rho}
\rho(k):=\frac12\sum_{\alpha\in\Delta_+}k_\alpha \alpha \, ,
\end{equation}
 we see that in addition to the usual half-sum of the positive roots
 $\rho=\rho(k_0)=\frac12\sum_{\alpha\in\Delta_+}\alpha=\sum_{i=1}^n\omega_i$ we have 
\begin{equation}\label{rhosl}
\rho^s:=\rho(k^s)=\frac12\sum_{\alpha\in\Delta^s_+}\alpha=\sum_{\alpha_i\in\Pi^s}\omega_i\,,\quad
\rho^l:=\rho(k^l)=\frac12\sum_{\alpha\in\Delta^l_+}\alpha=\sum_{\alpha_i\in\Pi^l}\omega_i\,.
\end{equation}

To $\rho^s$ and $\rho^l$ correspond the important short and long Coxeter numbers
$h^s$ and $h^l$ defined by
\begin{equation}\label{hshl}
h^s:=\langle \rho^s,\gamma^{s\vee}\rangle+1 \,,\qquad h^l:=\langle \rho^l,\gamma^{l\vee}\rangle+1\,.
\end{equation}
The explicit calculations using the values in Table \ref{rootnumbers} imply that
\begin{equation}\label{h-values}
h^s=1+\sum_{\alpha_i\in\Pi^s}m_i^\vee=\sum_{\alpha_i\in\Pi^s}m_i\,,\qquad
h^l=\sum_{\alpha_i\in\Pi^l}m_i^\vee=1+\sum_{\alpha_i\in\Pi^l}m_i\,.
\end{equation}

\section{$W-$invariant and $W-$skew invariant functions on $\mathbb{T}$}

\subsection{Sign homomorphisms}\ 

In addition to the usual sign homomorphisms on the Weyl group $W$ there
are two others. This is well known, but since it is short we prove it.
An abstract presentation determining $W$ is 
\[\langle r_1, \dots, r_n \,\mid r_i^2 =1, \, (r_ir_j)^{a_{ij}} =1, \, i,j = 1,\dots, n, \, i\ne j \rangle\,,\]
where $a_{ij} =2,3,4,6$ according as nodes $i$ and $j$ in the Coxeter-Dynkin diagram are not joined, joined by a single bond, a double bond, or a triple bond. 
Any homomorphism $\sigma: W \longrightarrow \{\pm 1\}$ is determined by the values on the generators $r_i$, $i=1,\dots, n$. The necessary and sufficient condition for $\sigma$ to be a homomorphism is that $(\sigma(r_i)\sigma(r_j))^{a_{ij}} =1$ for all $i\neq j$. This is automatically satisfied if $a_{ij}$ is even. When $a_{ij}$ is odd, i.e. $a_{ij} =3$, we need $\sigma(r_i) =\sigma(r_j)$.
Looking at the Coxeter-Dynkin diagrams we see that this allows precisely one choice of sign for all the short reflections 
and one for all the long reflections, and no other. Note that it does not matter whether or not we have a reflection in simple root or in any root since for any two roots $\alpha,\beta$ of the same length there exists $w\in W$ such that $r_\alpha=wr_\beta w^{-1}$ which implies $\sigma(r_\alpha)=\sigma(r_\beta)$.  Thus there are four homomorphisms $\sigma$:
\begin{equation}\label{homos}
\begin{alignedat}{3}
\mbox{id}\ &:&\quad&\text{all signs equal to}\, 1 \,
           &\quad &\text{(the trivial homomorphism)};\\
\det   &:&\quad  &\text{all signs equal to} \,-1\, 
          &\quad &\text{(the determinant)};\\
\sigma^l &:&\quad &\text{all long signs equal to} -1,
          &\quad &\text{all short signs equal to}\ 1 ;\\
\sigma^s &:&\quad &\text{all short signs equal to} -1,
          &\quad &\text{all long signs equal to} \ 1 .\\
\end{alignedat}
\end{equation}\
We shall use all four homomorphisms to introduce various classes of $W-$orbit functions. 

\subsection{$C,S,S^l-$ and $S^s-$functions}\

Let us fix the notation for the functions of the four families of $W$-orbit functions given by the homomorphisms \eqref{homos}. At first recall the definition of $C-$ and $S-$functions which were studied in \cite{KP06,KP07}.

\begin{equation}\label{CSfunctions}
C_\lambda(x)=\sum_{\mu\in O(\lambda)} e^{2\pi i\langle \mu,x\rangle}\,,\qquad
S_{\lambda+\rho}(x)=\sum_{w\in W} \det(w)  e^{2\pi i\langle w(\lambda+\rho),x\rangle} 
= \sum_{\mu \in O(\lambda+ \rho)} \sigma(\mu)  e^{2\pi i\langle \mu,x\rangle} \,.
\end{equation}
Here the parameter $\lambda\in P^+$ is a dominant weight, the variable 
$x\in\R^n$,  
$O(\lambda)$ is the $W$ orbit of $\lambda$, and $\sigma(\mu) := \sigma(w)$
where $\mu = w(\lambda + \rho)$.  Then
$|O(\lambda)|=\frac{|W|}{|\mbox{stab}_W\lambda|}$ 
is the number of points in $O(\lambda)$
where $|W|$ denotes the order of the Weyl group and $|\mbox{stab}_W\lambda|$ is the number of points in the stabilizer in $W$ of $\lambda$. For $S-$functions, the summation is in fact over the whole of $W$ since $\lambda+\rho$ has a trivial stabilizer. 

When there are two different root lengths there are two other orbit functions, arising from the homomorphisms $\sigma^s$ and $\sigma^l$:
\begin{equation}\label{functions}
S^s_{\lambda+\rho^s}(x)=\sum_{\mu\in O(\lambda+\rho^s)} \sigma^s(\mu)  e^{2\pi i\langle \mu,x\rangle}\,,\qquad
S^l_{\lambda+\rho^l}(x)=\sum_{\mu\in O(\lambda+\rho^l)} \sigma^l(\mu)  e^{2\pi i\langle \mu,x\rangle}\,,
\end{equation}
where $\rho^s,\rho^l$ are given by \eqref{rhosl}. Here again we are defining $\sigma^s(\mu):=\sigma^s(w)$ for $w\in W$ such that $\mu=w(\lambda+\rho^s)$ and $\sigma^l(\mu):=\sigma^l(w)$ for $w\in W$ such that $\mu=w(\lambda+\rho^l)$. 
This makes sense because the stabilizer in $W$ of $\rho^s$ is generated by long reflections $r_i$, so $\sigma^s$ takes the constant value $1$ on the stabilizer. Similarly,  $\sigma(\mu)$ in \eqref{CSfunctions} and $\sigma^l(\mu)$ are well defined.  

Evidently the $C$-functions are $W$ invariant while the $S$ (respectively $S^{s}$, $S^{l}$)-functions are $\det$ (respectively $\sigma^{s}$, $\sigma^{l}$)-skew invariant.

All of these functions can be viewed as functional forms of formal exponential sums
from $\C[P] $ of all linear combinations of formal exponentials $e^{\mu}$
with $\mu \in P$. In fact they are in $\Z[P]$ since all the coefficients are integers. 
We write $\C[P]^{W}$ (respectively $\C[P]^{s}$, $\C[P]^{l}$) for the $W$ invariant
(respectively $\sigma^{s}$, $\sigma^{l}$-skew invariant) exponential sums, and similarly for the
corresponding integral forms. More about the relationship between the formal exponentials
and their use as functions may be found in \cite{MPCubature}. 

The functions of $\C[P]$, as we have defined them are functions on 
$\R^{n}$. However, since they are periodic modulo $Q^{\vee}$, they may
be considered as functions on $\T \simeq \R^{n}/Q^{\vee}$. This is the way 
in which we shall normally think of them. For integration purposes, an integral
over $\T$ rewrites to an integral over a fundamental domain
for the lattice $Q^{\vee}$, for instance 
$\{\sum_{j=1}^{n} x_{j}\alpha^\vee_{j} \,:\, 0 \le x_{j} < 1 \;\mbox{for all} \; j \}$.

For notational convenience
we use 
\begin{equation}\label{defPhi}
\phi_{\mu} : x \mapsto e^{2\pi i \langle \mu,x\rangle} \, ,
\end{equation}
which for each weight $\mu \in P$ combines the exponential 
mapping $x \mapsto \exp(2 \pi i x)$ of $\mathfrak t$ to $\T$
and the $\C$-mapping $\exp(2 \pi i x) \mapsto e^{2\pi i \langle \mu, x\rangle}$
on $\T$. As we have just said, we may think of $\phi_{\mu}$ as a function on $\T$.

We note specially that the $S^s-$ and $S^l-$functions are sums over orbits rather than sums
over the entire Weyl group. Obviously they can be rewritten as Weyl group sums, but
in general these are redundant and 
for what follows the orbit sums are what we need. They also may be interpreted as functions
on $\T$ since they are invariant by $Q^{\vee}$-translations. 

\begin{tvr}\ \label{prod}
\begin{equation*}
S^s_{\rho^s}(x)=\Pi_{\alpha\in\Delta^s_+}\left(e^{\pi i\langle\alpha,x\rangle}-e^{-\pi i\langle\alpha,x\rangle}\right),\quad
S^l_{\rho^l}(x)=\Pi_{\alpha\in\Delta^l_+}\left(e^{\pi i\langle\alpha,x\rangle}-e^{-\pi i\langle\alpha,x\rangle}\right) \,.
\end{equation*}
\end{tvr}

\begin{proof}
We show the result for $S^s_{\rho^s}$, the proof for $S^l_{\rho^l}$ is similar.
Let $W^s$ denote the Weyl group generated by short reflections and $W^l$ the Weyl group generated by long reflections. Then $W$ can be written as a semi-direct product $W \simeq V^l \ltimes W^s$ where $V^l$ is a subgroup of $W^l$. We know that the stabilizer of $\rho^s$ is generated by long reflections, so $O(\rho^s)= W^s(\rho^{s})$ 
and
$$S^s_{\rho^s}(x)=\sum_{w\in W^s}\sigma^s(w)e^{2\pi i\langle w(\rho^s),x\rangle}\,.$$
Thus the result is simply the usual formula that holds for all root systems.
\end{proof}

We are especially interested in the {\bf hybrid-characters}:
\begin{equation}\label{hybrid}
\chi^l_\lambda(x)=\frac{S^l_{\lambda+\rho^l}(x)}{S^l_{\rho^l}(x)}\,,\qquad
\chi^s_\lambda(x)=\frac{S^s_{\lambda+\rho^s}(x)}{S^s_{\rho^s}(x)}\,.
\end{equation}
They are clearly $W$-invariant and we shall see that their linear span is $\C[P]^{W}$.
In particular they are well defined functions on all of $\mathfrak t$ (and, of course, they
can be considered as functions on $\T$).
The hybrid characters for the fundamental weights $\omega_{1}, \dots \omega_{n}$
also generate $\C[P]^{W}$ as a ring, and the main point is that they will become the new variables $X^s_{1}, \dots, X^s_{n}$ and $X^l_{1}, \dots, X^l_{n}$. In fact these hybrid characters are in $\Z[P]^{s}$ and $\Z[P]^{l}$ and what we just said applies
at the level of these rings.  These facts are well known, but because of their central importance
here we sketch out the proofs in what follows.

\begin{tvr}\label{denominatorFactor}
$\mathbb{Z}[P]^s=\mathbb{Z}[P]^W S_{\rho^s}^s$, \quad 
$\mathbb{Z}[P]^l=\mathbb{Z}[P]^W S_{\rho^l}^l$.
\end{tvr}
\begin{proof}
Inclusions in one direction are obvious.
We show the reverse inclusion in the short case. Let $f\in\mathbb{Z}[P]^s$ and write $f=\sum_{\mu\in P}c_{\mu}e^{\mu}$. Let $\alpha\in\Delta^s_+$. 
Then $-f=r_{\alpha}f=\sum c_{\mu}e^{r_{\alpha}\mu}$ and so, 
$-f=\sum -c_{r_\alpha\mu}e^{r_{\alpha}\mu}=\sum -c_{\mu}e^{\mu}=\sum c_{\mu}e^{r_{\alpha}\mu}$. 

Thus we can divide $\set{\mu}{c_{\mu}\neq0}$ into pairs $\{\mu_1,\mu_2\}$ where $\mu_2=r_{\alpha}\mu_1$, $c_{\mu_2}=-c_{\mu_1}$, and $\mu_{1} \preceq \mu_{2}$ 
(if $\mu_1=\mu_2$ then $c_\mu=-c_{\mu}$, so $c_{\mu}=0$). 
Thus $f=\sum_{\mu\in S}c_{\mu}\left(e^{\mu}-e^{\mu-z_\alpha\alpha}\right)$ 
for some finite subset $S\subset P$. 

Since $e^{\mu}-e^{\mu-z_\alpha\alpha}=e^{\mu}\left(1-e^{-z_\alpha\alpha}\right)$ and $\left(1-e^{-\alpha}\right)$ is always a factor of $\left(1-e^{-z_\alpha\alpha}\right)$, we obtain $f=\left(1-e^{-\alpha}\right)f_\alpha$ for some $f_\alpha \in \mathbb{Z}[P]$; 
and this statement is true for every $\alpha\in\Delta^s_+$. Now using \cite{B} Ch.6, we have that $\set{1-e^{- \alpha}}{\alpha\in\Delta_+}$ are all relatively prime, and hence from 
$\left(1-e^{-\alpha}\right)\mid f$ for each $\alpha\in\Delta_+^s$ we obtain $\Pi_{\alpha\in\Delta^s_+}\left(e^{\alpha/2}-e^{-\alpha/2}\right)\mid f$. The result now follows.
\end{proof}  
 
\subsection{Domains $F^s$ et $F^l$}\ \label{fund}

The $S^{s}$-functions are $\sigma^{s}$-skew invariant and are also translationally invariant
with respect to $Q^{\vee}$. As such they are determined entirely by their restriction
to the fundamental region $F$. 
Because of  Props. \ref{prod} and \ref{denominatorFactor}, the $S^{s}$-functions
vanish on the root hyperplanes of $F$ that correspond to the short roots, namely
on $H^s:=\bigcup_{\alpha_j\in\Pi^s} H_j$. Define $F^s:=F\setminus H^s$.
We shall be interested in the $S^{s}$-functions and their corresponding hybrid
characters on this new domain.

All this can be done for the $S^{l}$-functions too, and we define 
$H^l:=H_0\cup\bigcup_{\alpha_j\in\Pi^l}H_j$ and $F^l:=F\setminus H^l$.
Note that the hyperplane $H_{0}$ appears in this case, since it is always associated
with reflection in a long root. 

Using \eqref{funddom}, the domains $F^s$ and $F^l$ can be described by
\begin{equation}\label{slfund}
\begin{aligned}
F^s&=\set{y^s_1\omega_1^\vee+\dots +y^s_n\omega_n^\vee}{y_0^s+\sum_{i=1}^nm_iy_i^s=1\text{ and }y_i^s\in\R^{>0}\text{ if }\alpha_i\in\Pi^s \text{ otherwise } y_i^s\in\R^{\geq0}};\\
F^l&=\set{y^l_1\omega_1^\vee+\dots +y^l_n\omega_n^\vee}{y_0^l+\sum_{i=1}^n m_iy_i^l=1\text{ and }y_0^l,y_i^l\in\R^{>0}\text{ if }\alpha_i\in\Pi^l \text{ otherwise } y_i^l\in\R^{\geq0}}\,.
\end{aligned}
\end{equation}

Although $F^{s}$ and $F^{l}$ are proper subsets of $F$, it is more relevant that each of them
is a proper superset of $F^{\circ}$. The original domain $\Omega \subset \C^{n}$ arises
 as a continuous image of $F^{\circ}$ via the mapping $\Xi$ \eqref{polyMap}. The corresponding domains in the hybrid cases arise from
in a similar way from these two supersets:
\begin{equation}\label{sl-Domains}
\Omega^{s}:= \Xi^{s}(F^{s}) \supset \Omega
\qquad \Omega^{l}:= \Xi^{l}(F^{l}) \supset \Omega \,.
\end{equation}
These will appear when we switch from variables $x$ to variables $X$.

\subsection{Jacobi polynomials}\ \label{Jacobi}

All the characters $\chi_{\lambda}$, the hybrid characters $\chi^{s}_{\lambda}$,
$\chi^{l}_{\lambda}$,  and the $C$-functions $C_{\lambda}$, $\lambda \in P^{+}$ 
lie in $\Z[P]^{W}$. Furthermore each set
forms a $\Z$-basis for it and in each case the characters or hybrid characters indexed
by the fundamental weights $\omega_{j}$, $j=1, \dots, n$, generate $\Z[P]^{W}$
as a polynomial ring. Of course these facts apply to $\C[P]^{W}$ as well. 
This is quite easy to see because it
is obvious that the $C$-functions $C_{\lambda}$, $\lambda \in P^{+}$, 
are a $Z$-basis for $\Z[P]^{W}$ and the others can be written as 
sums of the form 
\[C_{\lambda} + \sum_{\setcomb{\phantom{ii}\mu \in P^{+}}{\mu \prec \lambda} } a_{\lambda, \mu}C_{\mu}\]
where the $a_{\lambda, \mu}\in \Z$. This triangular form with unit diagonal coefficients
can be inverted in $\Z[P]^{W}$, showing that each of the other sets is a basis too.
Similarly each $C_{\lambda}= C_{k_{1}\omega_{1}+ \cdots + k_{n}\omega_{n}} $ 
can be written in the form
\[C_{\omega_{1}}^{k_{1}} \cdots   C_{\omega_{n}}^{k_{n}}  + \sum_{\setcomb{\phantom{ii}\mu \in P^{+}}{\mu \prec \lambda} } a_{\lambda, \mu}C_{\mu}\]
with integer coefficients, and this provides the recursive step to write any element of $\Z[P]^{W}$
as a polynomial in the $C_{\omega_{j}}$. The same thing can be done with the fundamental
characters or hybrid characters. 

Although we have no need for the specific values of the coefficients in these
expressions, there are ways to compute them. As a specific example there are the Jacobi polynomials $P(\lambda,k)$, defined
for any multiplicity function $k$, see \cite{hec}, and any $\lambda\in P^+$ by 
\begin{equation}\label{jacobi}
P(\lambda,k)=\sum_{\setcomb{\phantom{ii}\mu \in P^{+}}{\mu \preceq \lambda}}c_{\lambda\mu}(k)C_\mu,\end{equation}
where the coefficients $c_{\lambda\mu}(k)$ are defined recursively by:
\begin{equation}\label{coef}
{(\lambda+\rho(k)\mid\lambda+\rho(k))-(\mu+\rho(k)\mid\mu+\rho(k))}c_{\lambda\mu}(k)=2\sum_{\alpha\in\Delta_+}k_\alpha\sum_{j=1}^\infty(\mu+j\alpha\mid\alpha)c_{\lambda,\mu+j\alpha}
\end{equation}
along with the initial value $c_{\lambda\lambda}=1$ and the assumption 
$c_{\lambda\mu}=c_{\lambda,w(\mu)}$ for all $w\in W$.
 Recall that $\rho(k)$ is given by \eqref{rho}. 
 
 For $k=k_0$ this relation \eqref{coef} is the Freudenthal recurrence relation used to find the coefficients of decomposition of characters $\chi_{\lambda}=\frac{S_{\lambda+\rho}}{S_\rho}$ of irreducible representations of simple Lie algebras into $C-$functions. In other words,
\begin{equation}\label{char}
\chi_{\lambda}=P(\lambda,k_0)=\sum_{\setcomb{\phantom{ii}\mu \in P^{+}}{\mu \preceq \lambda}}c_{\lambda\mu}(k_0)C_\mu\,.
\end{equation}
Furthermore \eqref{jacobi}, for $k^s$ and $k^l$ be given by \eqref{kfunction} 
and $\lambda\in P^+$, we have
\begin{equation*}
\chi_\lambda^s=P(\lambda,k^s) \quad \text{and}\quad \chi_\lambda^l=P(\lambda,k^l)\,.
\end{equation*}

\subsection{An inner product on $\C[P]^W$}\label{inp}\

The standard inner product on $\C[P]$ is defined by
\begin{equation}\label{in}
\langle f,g\rangle_{\mathbb{T}}=\int_{\mathbb{T}}f\overline{g}d\theta_{\mathbb{T}}\,,
\end{equation}
where $d\theta_{\mathbb{T}}$ is the normalized Haar measure on the torus $\mathbb{T}$. Relative to this, the functions $\phi_{\lambda}$ \eqref{defPhi} form an orthogonal basis of $\C[P]$. Its completion is
the Hilbert space $L^2(\mathbb{T},\theta_{\mathbb{T}})$. We let $L^2(\mathbb{T},\theta_{\mathbb{T}})^W$ be the subspace of all W-invariant elements of $L^2(\mathbb{T},\theta_{\mathbb{T}})$, which is in fact the closure of $\C[P]^W$ in $L^2(\mathbb{T},\theta_{\mathbb{T}})$.

We now modify this inner product in a natural way so that the hybrid-characters $\chi^s_{\lambda}$ (or $\chi^l_{\lambda}$) form an orthogonal basis for $L^2(\mathbb{T},\theta_{\mathbb{T}})^W$. 
Notice here that we are interpreting functions as functions on $\T$.

For any element $f\in L^2(\mathbb{T},\theta_{\mathbb{T}})^W$, we have $fS_{\rho^s}^s\in L^2(\mathbb{T},\theta_{\mathbb{T}})$. One can form its Fourier expansion
\[fS^s_{\rho^s}=\sum_{\mu\in P}\langle fS^s_{\rho^s},\phi_\mu\rangle_\mathbb{T}\phi_\mu\,,\]
and since $fS^s_{\rho^s}$ is $\sigma^{s}$-skew-invariant with respect to $W$, this can be rewritten as
\[fS^s_{\rho^s}=\sum_{\lambda\in P^+}\langle fS^s_{\rho^s},\phi_{\lambda+\rho^s}\rangle_{\mathbb{T}}\sum_{\mu'\in O(\lambda+\rho^s)}\sigma^s(\mu')\phi_{\mu'}=\sum_{\lambda\in P^+}\langle fS^s_{\rho^s},\phi_{\lambda+\rho^s}\rangle_{\mathbb{T}} S^s_{\lambda+\rho^s}\,.\]
Dividing by $S^s_{\rho^s}$ we have
\[f=\sum_{\lambda\in P^+}\langle fS_{\rho^s}^s,\phi_{\lambda+\rho^s}\rangle_{\mathbb{T}}\,\chi^s_{\lambda}\,,\]
and then by the $W-$invariance of $\theta_{\mathbb{T}}$ and $\sigma^{s}$-skew-invariance of $fS_{\rho^s}^s$, we obtain
\begin{equation*}
\begin{aligned}
\langle fS^s_{\rho^s},\phi_{\lambda+\rho^s}\rangle_\mathbb{T}&=\int_\mathbb{T}fS^s_{\rho^s}\overline{\phi_{\lambda+\rho^s}}d\theta_\mathbb{T}=
\frac{1}{|W|}\int_\mathbb{T}\sum_{w\in W}\sigma^s(w)fS^s_{\rho^s}\overline{\phi_{w(\lambda+\rho^s)}}d\theta_{\mathbb{T}}\\&=\frac{|\mbox{stab}_W(\lambda+\rho^s)|}{|W|}\int_{\mathbb{T}}fS^s_{\rho^s}\overline{S_{\lambda+\rho^s}^s}d\theta_\mathbb{T}=|\mbox{stab}_W(\lambda+\rho^s)|\int_{F^s}f\overline{\chi_\lambda^s}S^s_{\rho^s}\overline{S_{\rho^s}^s}d\theta_\mathbb{T}\,.
\end{aligned}
\end{equation*}

This suggests the new inner product on $L^2(\mathbb{T},\theta_{\mathbb{T}})^W$ as
\begin{equation*}
(f,g)_s=\int_{F^s}f\overline{g}S^s_{\rho^s}\overline{S^s_{\rho^s}}d\theta_{\mathbb{T}}.
\end{equation*}
Then, we can write
\begin{equation}\label{s-TypeFourierExpansion}
f=\sum_{\lambda\in P^+}|\mbox{stab}_W(\lambda+\rho^s)|(f,\chi^s_\lambda)_s\chi^s_\lambda\,.
\end{equation} 
In particular, with $f= \chi^{s}_{\mu}$
we have
\begin{equation}\label{s-TypeOrthogonality}
\chi^{s}_{\mu}=\sum_{\lambda\in P^+}|\mbox{stab}_W(\lambda+\rho^s)|
(\chi^{s}_{\mu},\chi^s_\lambda)_s\chi^s_\lambda
\,,
\end{equation}
from which we have the orthogonality relations
\begin{equation}\label{s-Ortho}
(\chi^{s}_{\mu},\chi^{s}_{\lambda})_{s} = 
\frac{1}{|\mbox{stab}_W(\mu+\rho^s)|} \delta_{\mu\lambda}
\end{equation}
Writing this out, we have 
\begin{tvr}\label{s-cntsOrthogonality}
For $\lambda,\mu\in P^+$, 
\begin{equation*}
\int_{F^{s}} S^s_{\lambda+\rho^s}(x)
\overline{S^s_{\mu+\rho^s}(x)}\,d\theta_{\T}(x)=
(\chi^{s}_{\lambda},\chi^{s}_{\mu})_{s}=\frac{1}{|\mbox{stab}_W(\lambda+\rho^s)|}\delta_{\lambda\mu}\,;
\end{equation*}
where  $|\mbox{stab}_W(\lambda+\rho^s)|$ denotes the number of elements in stabilizer of $\lambda +\rho^s$ in $W$.
The parallel result holds for the long root case. 
\end{tvr}

\section{Polynomial variables and elements of finite order}\

The cubature formulas rely on being able to identify the ring
$\C[x_{1}, \dots, x_{n}]^{W}$ as a polynomial ring and then forming the connection
between the variables $X_{j}$ and characters on $\G$ (treated as 
functions on $\mathfrak t$). In the usual case, the characters
are the characters of the fundamental representations with highest weight $\omega_{j}$.
In the hybrid cases we use hybrid characters instead.  As we shall see, they all generate
essentially the same ring, but the explicit mappings between the natural variables of $\mathfrak t$
and the variables $X_{j}$ are different. We shall work specifically with the short case, the long case being in every way parallel to it.

\subsection{Polynomial variables for the hybrid cases}\

Let $X_1^s,\dots, X^s_n$ denote the polynomial variables defined by
\begin{equation*}
 X_1^s:=\chi_{\omega_1}^s(x),\dots, X^s_n:=\chi_{\omega_n}^s(x),\quad x\in F^s\,,
\end{equation*}
where $\chi^s_{\omega_j}$ are the fundamental hybrid-characters \eqref{hybrid}.

As in \cite{MPCubature} , we define the $m-$degree of the variables $X_1^s,\dots, X^s_n$ by 
assigning degree $m_i^\vee$ to $X^s_i$. Thus
the monomial $(X_1^s)^{\lambda_1}\dots (X_n^s)^{\lambda_n}$ has $m$-degree
$\lambda_1 m_1^\vee+\dots+\lambda_n m^\vee_n$ and 
the dimension of the space of
polynomials of $m$-degree at most $M$ is the cardinality of the set
\begin{equation}\label{spaceDimension}
\set{(\lambda_1,\dots,\lambda_n)}{\sum_{i=1}^n m^{\vee}_i\lambda_i\leq M,\lambda_i\in\Z^{\geq0}}\,.
\end{equation}

In addition, we say that $\lambda=(\lambda_1,\dots,\lambda_n)=\lambda_1\omega_1+\dots+\lambda_n\omega_n$ has $m-${\bf degree} equal to 
\begin{equation}\label{lamdeg}
\langle\lambda,\gamma^{s\vee}\rangle=\lambda_1m_1^\vee+\dots+\lambda_n m_n^\vee\,.
\end{equation}

The new variables give rise to the mapping
\[ \Xi^{s}:  x \mapsto (X^{s}_{1}(x), \dots, X^{s}_{n}(x)) \in \C^{n},  \quad x\in F^{s}
\]
and similarly $\Xi^{l}$. These mappings are injective since the values
of these fundamental hybrid characters determine the values of all the characters (hybrid or otherwise), hence a specific conjugacy class in $\G$, and finally, then, a unique
point in $F$. Then we have the domain
\begin{equation*}
\Omega^s:= \Xi^s(F^{s}) = \set{(X_1^s(x),\dots,X_n^s(x))}{x\in F^s}\,.
\end{equation*}
Evidently this is a subset of $\C^{n}$, but in fact $\Omega^s \subset\R^n$. By \S\ref{Jacobi}, we see that each variable $X_i^s$ can be written as a polynomial in fundamental characters 
$\chi_{\omega_k}$ with integer coefficients. As discussed in \cite{MPCubature}, we know that $\overline{\chi_{\omega_k}}=\chi_{\omega_k}$ for algebras with two roots lengths. Therefore, we also have $\overline{X_i^s}=X_i^s$ and thus $\Omega^s\subset \R^n$.

We define
\begin{equation}\label{K}
K^{s}:=\frac{S^s_{\rho^s}\overline{S^s_{\rho^s}}}{S^l_{\rho^l}\overline{S^l_{\rho^l}}}.
\end{equation}
This function arises as a kernel in the integral of the cubature formulas for the short root case. 
The denominator of $K^{s}$ does not vanish anywhere on the interior $F^\circ$ of the fundamental domain $F$, so 
$K^{s}$ is defined on this region. 
$K^{s}$ is a $W$-invariant rational function and can be rewritten as a function in terms of the fundamental hybrid-characters $\chi_i^s$. We can regard $K^{s}$ as a strictly positive function on $F^\circ$ or as a function in the variables $X^s_i$ on the interior of $\Omega^s$,
\begin{equation*}
K^{s}=K^{s}(X^s_1,\dots,X^s_n)=\frac{S^s_{\rho^s}(x)\overline{S^s_{\rho^s}(x)}}{S^l_{\rho^l}(x)\overline{S^l_{\rho^l}(x)}}\,,\qquad x\in F^\circ\,,
\end{equation*}
see Remark~\ref{xXremark}

Along with $K^{s}$ we define $\kappa^{s}$ on $\Omega^{s}$ by 
\begin{equation}\label{kappa}
\kappa^{s}(X^{s}) =
\frac{|Wx|}{|W|}S^s_{\rho^s}(x)\overline{S^s_{\rho^s}(x)} \,.
\end{equation}
Note that $|Wx|$ is just a number of points in $W$-orbit of $x$ in $\mathfrak{t}/Q^\vee$ and its value is uniquely
associated with $X^{s}$ since $\Xi^{s}$ is injective.

The $S^l$-functions are handled in the same way. Just interchange $s$ and $l$ in the discussion 
above. In particular notice that $K^{l}= (K^{s})^{-1}$ on $F^{\circ}$. We emphasize here that the $m$-degree of $\lambda$ for the long root case is equal to 
\begin{equation}
\langle\lambda,\gamma^{l\vee}\rangle=\lambda_1m_1^{l\vee}+\dots+\lambda_n m_n^{l\vee}\,.
\end{equation}
and thus is not the same as $m$-degree \eqref{lamdeg} of $\lambda$ for the short root case or the same as the $m$-degree of polynomials (see Table \ref{rootnumbers}).

\subsection{The Jacobian}\label{der}\

Although the cubature formulas that we are aiming to prove are set within the context
of the polynomial ring $\C[X_{1}, \dots, X_{n}]$, what underlies them is the realization
of the variables $X_{j}$ as functions, actually characters $\chi_{j}$ (or hybrid characters
$\chi_{j}^{s},\chi_{j}^{l}$), on $\T$.
These characters are first
of all functions on $\T$, but are treated also as functions on $\mathfrak t$ via the exponential
map -- indeed they are exponential sums. As functions on $\mathfrak t$ they become
functions of $n$ variables in terms of the standard basis $\{\alpha_{1}^{\vee}, \dots,
\alpha_{n}^{\vee}\}$. In order to make transitions from the $\alpha^{\vee}$-variables
to the $X_{j}$-variables we require the Jacobian $J$ with matrix entries  
$J_{jk}=D_{\alpha^\vee_j}\chi_{\omega_k}$, see definition below. This is written for the case of the 
characters, and in this case the Jacobian was determined in \cite{MPCubature}.
Since the transition from characters to hybrid characters is made through a unipotent
transformation, the determinant of the Jacobian is not altered for the hybrid characters.
\begin{tvr}\label{steinberg}
\begin{equation}
\det(J) = \det(J^s)=\det(J^l)=S_\rho=S_{\rho^s}^sS^l_{\rho^l}\,.
\end{equation}
\end{tvr}

Note that from this we have 
\begin{equation}\label{KJrelation}
(K^{s})^{1/2} |\det(J)| = |S^{s}_{\rho_{s}}|^{2} \,.
\end{equation}

With $x=(x_1,\dots,x_n)=x_1\alpha_1^\vee+\dots+x_n\alpha_n^\vee$
as variables on $\mathfrak t$ and the derivation mapping $D_{\alpha_i^\vee}$ defined by
\begin{equation*}
D_{\alpha_j^\vee}e^{\langle\lambda,2\pi i x\rangle}=\langle\lambda,\alpha_j^\vee\rangle e^{\langle\lambda,2\pi i \sum_{k=1}^nx_k\alpha^\vee_k\rangle}=\frac{1}{2\pi i}\frac{d}{dx_j}e^{\langle\lambda,2\pi ix\rangle}\,,
\end{equation*}
we compute 
\[D_{\alpha_j^\vee} \chi^{s}_{\omega_{k}} 
=\frac{1}{2 \pi i} \frac{d}{d x_{j}} \chi^{s}_{\omega_{k}}  \,.\] 
Then proposition \ref{steinberg} implies that the Jacobian of the transformation from the variables $x$ to variables $X^s$ or $X^l$ is
\begin{equation*}
|(2\pi i)^nS_\rho(x)|=(2\pi)^n|S_\rho(x)|\,.
\end{equation*}
So, by \eqref{KJrelation}, we have 
\begin{equation}\label{sint}
\begin{aligned}
\int_{\Omega^s}f\overline{g}(K^s)^{1/2}\,dX^{s} &=\int_{\Omega^s}f(X_1^s,\dots,X_n^s)\overline{g(X_1^s,\dots,X_n^s)}(K^s)^{1/2}(X_1^s,\dots,X_n^s) \, dX_1^s\dots dX_n^s\\
&=(2\pi)^n\int_{F^s}f(x)\overline{g(x)}S^s_{\rho^s}(x)\overline{S^s_{\rho^s}(x)}\,dx\,.
\end{aligned}
\end{equation}
Particularly note the special case of this when $f= \chi^{s}_{\lambda}$ and 
$g=\chi^{s}_{\mu}$ when, along with Prop.~\ref{s-cntsOrthogonality}, it becomes
\begin{equation}\label{transition}
 (2\pi)^{-n}\int_{\Omega^s}\chi^{s}_{\lambda}\overline{\chi^{s}_{\mu}}(K^s)^{1/2}\,dX
=\int_{F^s} S^s_{\lambda+ \rho^s}(x)\overline{S^s_{\mu+ \rho^s}(x)}\,dx 
= (\chi_{\lambda},\chi_{\mu})_{s}\,.
\end{equation}

Note that the integrals over $\Omega^s$ are well defined since $(K^{s})^{1/2}\, dX^{s}$ is defined over the interior of $\Omega^s$ and is zero on its boundary.

\subsection{Cones of elements of finite order}\

Every conjugacy class of elements of $G$ meets the fundamental chamber in $\T$
and so is $\exp{2 \pi i x}$ a unique $x\in F$. The elements of finite order (EFO) are particularly
interesting because they provide a way of discretization that is intuitive, natural, 
and computationally efficient. The conjugacy classes of elements of finite order $N$ (this includes
all elements whose order divides $N$) are precisely
given by $\frac{1}{N} Q^{\vee} \cap F$ and those of adjoint order $N$,
i.e. of order $N$ in the adjoint representation of $G$ on itself, are given by
$\frac{1}{N} P^{\vee} \cap F$ \cite{MP84}. It is these latter elements that will define the nodes for
the cubature formula. More particularly, having chosen some positive integer $M$, 
we wish to use 
\begin{equation*}
F^s_{M+h^s}:=\frac{1}{M+h^s}P^\vee\cap F^s\,,\qquad F^l_{M+h^l}:=\frac{1}{M+h^l}P^\vee\cap F^l \,,
\end{equation*}
where $h^s,h^l$ are defined by \eqref{hshl}. Using \eqref{slfund}, the elements of the fragments can be represented as follows.
\begin{equation}\label{grid1}
\begin{aligned}
x\in F^s_{M+h^s}\quad\Longleftrightarrow\quad &x=\frac{1}{M+h^s}(s^s_1\omega_1^\vee+\dots + s^s_n\omega_n^\vee)\text{ with the coordinates $(s^s_1,\dots,s^s_n)$ satisfying }\\
&s_0^s+\sum_{i=1}^nm_is_i^s=M+h^s\,,\text{ where }s_i^s\in\N\text{ if }\alpha_i\in\Pi^s \text{ otherwise } s_i^s\in\Z^{\geq0}\,;
\end{aligned}
\end{equation} 
\begin{equation}\label{grid2}
\begin{aligned}
x\in F^l_{M+h^l}\quad\Longleftrightarrow\quad &x=\frac{1}{M+h^l}(s^l_1\omega_1^\vee+\dots + s^l_n\omega_n^\vee)\text{ with the coordinates $(s^l_1,\dots,s^l_n)$ satisfying }\\
&s_0^l+\sum_{i=1}^nm_is_i^l=M+h^l\,,\text{ where }s_0^l,s_i^l\in\N\text{ if }\alpha_i\in\Pi^l \text{ otherwise } s_i^l\in\Z^{\geq0}\,;
\end{aligned}
\end{equation} 
The coordinates $[s_0^s,s_1^s,\dots,s_n^s]$ and $[s_0^l,s_1^l,\dots,s_n^l]$ are called the Kac coordinates of $x$, \cite{MP84}.
 
Since $h^{s}=\sum_{\alpha_i\in\Pi^s}m_i$ \, and \, $h^l=1+\sum_{\alpha_i\in\Pi^l}m_i$, 
each of the sets $F_{M+h^s}$ and $F_{M+h^l}$ has the same cardinality as the set:
\begin{equation}\label{set}
\set{(t_1,\dots,t_n)}{\sum_{i=1}^nm_it_i\leq M,t_i\in\Z^{\geq0}}\,.
\end{equation}
 The explicit formulas for the cardinality of $F^s_{M+h^s}$ and $F^l_{M+h^l}$ have been calculated for all $M$ and for all simple Lie algebras in \cite{HMP}.
 
 Comparing \eqref{spaceDimension} and \eqref{set}, and using the 
 the fact that the marks and co-marks are just permutations of each other (see
 Table~\ref{rootnumbers}), we see the important fact:
\begin{thm}
The number of monomials in $\C[X^s_1,\dots,X^s_n]$ of $m-$degree at most $M$ is equal to the number of points in $F^s_{M+h^s}$. The parallel result holds for long root case.
\end{thm}

\subsection{Points of $F_{M+h^s}^s$ as zeros of $S^s-$functions}\label{comzeros}\

It is very interesting that the points that will be the nodes for the cubature formulas
are also distinguished by being zeros of certain $S^{s}$-functions.

\begin{tvr} \label{zerosThm} Let $M\in\Z^{\geq0}$.
The functions $S^s_{\lambda+\rho^s}$ and the hybrid-characters 
$\chi_{\lambda}^s$ with $\lambda$ of $m-$degree $= M+1$ vanish 
at all points of $F^s_{M+h^s}$.  
The same is true with $s$ replaced by $l$ throughout. 
\end{tvr} 
\begin{proof}
We denote by $r$ the reflection in the highest short root $\gamma^{s}$, on the root and co-root side, given respectively by
\begin{equation*}
\begin{aligned}
&r(\lambda):=r\lambda=\lambda-\langle\lambda,\gamma^{s\vee}\rangle\gamma^s\,,\\
&r(x):=rx=x-\langle\gamma^s,x\rangle\gamma^{s\vee}\,.
\end{aligned}
\end{equation*}
Let $\lambda=\lambda_1\omega_1+\dots+\lambda_n\omega_n\in P^+$. Divide the orbit $O=O(\lambda+\rho^s)$ into $O_+$ on which $\sigma^s$ takes the value $1$, and $O_-$ on which it  takes value $-1$, and note that $O_-=rO_+$. Then we can write
\[S^s_{ \lambda + \rho^s}(x)= \sum_{\mu\in O_+} 
(e^{2  \pi i \langle \mu, x\rangle} -  e^{2  \pi i \langle r\mu, x\rangle}) 
= \sum_{\mu\in O_+} 
(e^{2  \pi i \langle \mu, x\rangle} -  e^{2  \pi i \langle \mu, rx\rangle}) 
\,.\]
 Now, $S^s_{ \lambda + \rho^s}(x)$
 will vanish for all $x\in F^s_{M+h^s}$ if each term
\[ e^{2  \pi i \langle \mu, x\rangle} -  e^{2  \pi i \langle \mu, rx\rangle} =0 \, ,\]
or equivalently
\[\langle \mu, x\rangle -  \langle \mu, rx\rangle \in \Z \]
for all $x \in \frac{1}{M + h^s}P^\vee$.  
Since $x \in \frac{1}{M + h^s}P^\vee$ is $W$-invariant, this 
amounts to 
\begin{equation}\label{cond}\langle \lambda+ \rho^s, x\rangle -  \langle \lambda+ \rho^s, rx\rangle \in \Z \qquad\text{ for all } x \in \frac{1}{M + h^s}P^\vee\,,
\end{equation}
or equivalently
\begin{equation*}
\langle \gamma^s, x\rangle\langle \lambda+ \rho^s, \gamma^{s\vee}\rangle    \in \Z \qquad\text{ for all } x \in \frac{1}{M + h^s}P^\vee\,.
\end{equation*} 
Since $\langle \gamma^s, P^\vee\rangle\subset\Z$, we have $\langle \gamma^s, x\rangle\in \frac{1}{M+h^s}\Z$, and
it is sufficient that $\langle\lambda+ \rho^s, \gamma^{s\vee}\rangle\in(M+h^s)\Z$. 
Requiring $\langle\lambda+ \rho^s, \gamma^{s\vee}\rangle=M + h^s$
leads to the condition
\[ \langle \lambda,{\gamma^s}^\vee \rangle = M +1 \,.\]
by definition of $h^s$ \eqref{hshl}. This is the condition of the hypothesis of the proposition
and proves the result for the $S^{s}$-functions. 

To get to the characters $\chi^{s}$ we have
to divide by $S^{s}_{\rho^{s}}$. The latter vanishes only on the walls of $H^{s}$ and
these are not part of $F^{s}$, and so this division does not affect the outcome. 
 
The proof for the long root case is parallel. 
\end{proof}

Recall that $|\frac{1}{M+h^s}P^{\vee}/Q^{\vee}| = c_{\mathfrak g}
(M+h^s)^{n}$, where $c_{\mathfrak g}$ is the determinant of $C$ (which is
the value of the index $[P^{\vee}:Q^{\vee}]$). Of course
there is a parallel formula for the long root case.

\subsection{Discrete orthogonality of $S^s-$ and $S^l-$functions}\

\begin{tvr}\label{s-discOrthogonality}
Let $M\in\Z^{\geq0}$ and $\lambda,\mu\in P^+$  and suppose that
\[\mbox{for all} \; w,w'\in W, \quad
w(\lambda+\rho^s)-w'(\mu+\rho^s)\notin (M+h^s)Q \]
unless $\lambda=\mu$ and $w(\lambda+\rho^s)=w'(\lambda+\rho^s)$. Then 
\begin{equation}\label{sum}
\frac{1}{c_\mathfrak{g}|W|(M+h^s)^n }\sum_{x\in F^s_{M+h^s}}|Wx|S^s_{\lambda+\rho^s}(x)\overline{S^s_{\mu+\rho^s}(x)}=\frac{1}{|\mbox{stab}_W(\lambda+\rho^s)|} \delta_{\lambda\mu}\,.
\end{equation}
The parallel result holds for the long root case. We recall that $Wx$ is the $W$-orbit of $x$ in $\mathfrak{t}/Q^\vee$.
\end{tvr}

\begin{proof}
The summands appearing in \eqref{sum} are dependent only on the values of $x \mod{Q^\vee}$, so we can reduce$\mod{Q^\vee}$ (see Remark \ref{xXremark}). The set $F^s_{M+h^s}$ is mapped faithfully by the $S^{s}$-functions in this process.

We begin by replacing the sum over $F^s_{M+h^s}$ by a sum over the {\em group}
$\frac{1}{M+h^s}P^\vee/Q^\vee$. For each representative element $x\in F^s_{M+h^s}$
we can form its $W$-orbit $Wx$. If we had all of $\frac{1}{M+h^s}P^\vee\cap F$ we would get all of $\frac{1}{M+h^s}P^\vee/Q^\vee$. As  it is, we are missing the orbits of points $F\setminus F^s$ and these are all in $H^s$ on which the $S^s$-functions vanish. So we can add them without changing anything.
Thus 
\[\sum_{x\in F^s_{M+h^s}}|Wx|\,S^s_{\lambda+\rho^s}(x)\overline{S^s_{\mu+\rho^s}(x)}
= \sum_{x\in \frac{1}{M+h^s}P^\vee/Q^\vee} S^s_{\lambda+\rho^s}(x)\overline{S^s_{\mu+\rho^s}(x)} \,.\]
The two $S^{s}$ terms when expanded are sums of exponential functions $\exp(2\pi i \langle
\nu, x\rangle)$ (which are well defined as functions on $\frac{1}{M+h^s}P^\vee/Q^\vee$), where each $\nu$ is of the form $\nu=w(\lambda+\rho^s)-w'(\mu+\rho^s)$. Fixing $\nu$ and 
and summing over $x$, we get their sum over the group is zero as long as $\langle\nu,x\rangle \notin \Z$ for at least one $x$. This requirement is just the same as saying $\nu\notin (M+h^s)Q$. In view of our hypothesis this fails only if $\lambda=\mu$ and $w(\lambda+\rho^s)=w'(\lambda+\rho^s)$. In that case the sum is $c_{\mathfrak{g}}(M+h^s)^n$. This happens once for each element in $O(\lambda+\rho^s)$. Since $|O(\lambda+\rho^s)|=|W|/|\mathrm{stab}_W(\lambda+\rho^s)|$, we are done.
\end{proof}

For a slight different point of view on discrete orthogonality,
as well as an algorithm for calculation of $|Wx|$, see \cite{HMP}. 

\section{Integration formulas}

Our aim is to create cubature formulas for the integrals of the form
\begin{equation*}
\int_{\Omega^s}f^s\overline{g^s}(K^{s})^{1/2}dX_1^s\dots dX_n^s\,,\qquad \int_{\Omega^l}f^l\overline{g^l}(K^{l})^{1/2}dX_1^l\dots dX_n^l \,.
\end{equation*}
where $f^s,g^s$ are functions in the variables $X_1^s,\dots,X_n^s$ defined on $\Omega^s$ and $f^l,g^l$ are functions in the variables $X_1^l,\dots,X_n^l$ defined on $\Omega^l$. 
These cubature formulas depend on the two orthogonality results that
we have shown, namely Prop.~\ref{s-cntsOrthogonality} and Prop.~\ref{s-discOrthogonality}, the first
involving an integral over $\Omega^{s}$ and the second a finite sum over $F^s_{M+h^s}$,
which yield identical results. The discrete orthogonality relations require specific separation hypotheses
on the weights, so to make use of the equalities we need only to guarantee that these hold.
The same applies to the long root case too. The image of $F^s_{M+h^s}$ in $\Omega^{s}$ under 
$\Xi^{s}$ is written as $\mathcal{F}^s_{M+h^s}$, and similarly for the long root case.

\subsection{The key integration formulas}\

\begin{thm}\label{thmformule}\
\begin{enumerate}
\item[(i)]
Let $M\in\Z^{\geq0}$ and $f,g$ be any polynomials in $\C[X_1^s,\dots,X_n^s]$ with $m-deg(f)\leq M+1$ and $m-deg(g)\leq M$. Then
\begin{equation}\label{intformule}
\begin{aligned}
\int_{\Omega^s}&f\overline{g}K^{1/2}dX^s = \int_{\Omega^s}f\overline{g}K^{1/2}dX_1^s\dots dX_n^s\\
&=(2\pi)^n\int_{F^s}f(\chi^s_{\omega_1}(x),\dots,\chi^s_{\omega_n}(x))\overline{g(\chi^s_{\omega_1}(x),\dots,\chi^s_{\omega_n}(x))}S^s_{\rho^s}(x)\overline{S^s_{\rho^s}}(x)dx_1\dots dx_n\\
&=\frac{1}{c_\mathfrak{g}|W|}\left(\frac{2\pi}{M+h^s}\right)^n\sum_{x\in F^s_{M+h^s}}f(\chi^s_{\omega_1}(x),\dots,\chi^s_{\omega_n}(x))\overline{g(\chi^s_{\omega_1}(x),\dots,\chi^s_{\omega_n}(x))}|Wx| S^s_{\rho^s}(x)\overline{S^s_{\rho^s}}(x)\\
&=\frac{1}{c_\mathfrak{g}}\left(\frac{2\pi}{M+h^s}\right)^n
\sum_{(X_1^s,\dots, X_n^s)\in \mathcal{F}^s_{M+h^s}}f(X_1^s,\dots, X_n^s)
\overline{g(X_1^s,\dots, X_n^s)}\kappa^{s}(X_1^s,\dots, X_n^s) \\
&=\frac{1}{c_\mathfrak{g}}\left(\frac{2\pi}{M+h^s}\right)^n
\sum_{X^s\in \mathcal{F}^s_{M+h^s}}f(X^{s})
\overline{g(X^s)}\kappa^{s}(X^s) \,.
\end{aligned}
\end{equation}
\item[(ii)]
Let $M\in\N$ and $f,g$ be any polynomials in $\C[X_1^l,\dots,X_n^l]$ with $m-deg(f)\leq M$ and $m-deg(g)\leq M-1$. Then 
\begin{equation}
\int_{\Omega^l}f\overline{g}(K^l)^{1/2}dX^l = \frac{1}{c_\mathfrak{g}}\left(\frac{2\pi}{M+h^l}\right)^n \sum_{X^l\in \mathcal{F}^l_{M+h^l}}f(X^{l})
\overline{g(X^l)}\kappa^{l}(X^l) \,.
\end{equation}
\end{enumerate}
\end{thm}

\begin{proof}
By linearity of \eqref{intformule}, we can only consider the monomials 
$$(\chi^s_{\omega_1})^{\nu_1},\dots,(\chi^s_{\omega_n})^{\nu_n}\,,\qquad \text{where }\nu_1m_1^\vee+\dots+\nu_nm_n^\vee\leq N$$
with $N=M+1$ for $f$ and $N=M$ for $g$.

By Section \ref{Jacobi}, we see that such monomial decomposes as a linear combination of $\chi^s_\lambda$ with $\lambda \preceq \nu$ (see \S\ref{simpleLie}) and the coefficient of $\chi^s_\nu$ is equal to $1$.
 
Thus it is sufficient to prove that
\begin{equation*}
\int_{F^s}\chi_{\lambda}^s(x)\overline{\chi_{\mu}^s(x)}S^s_{\rho^s}(x)\overline{S^s_{\rho^s}}(x)dx=\frac{1}{c_{\mathfrak{g}}|W|(M+h^s)^n}\sum_{x\in F^s_{M+h^s}}\chi_{\lambda}^s(x)\overline{\chi_{\mu}^s(x)}|Wx| S^s_{\rho^s}(x)\overline{S^s_{\rho^s}}(x)\,
\end{equation*}
for $\lambda,\mu\in P^+$ such that $m-deg(\lambda)\leq M+1$ and $m-deg(\mu)\leq M$.
This is true from Prop.~\ref{s-cntsOrthogonality} and Prop.~\ref{s-discOrthogonality}, provided
the weight separation conditions of Prop.~\ref{s-discOrthogonality} apply, that is,
whenever $\lambda\neq \mu$, it never happens that $w(\lambda+\rho^s)-w'(\mu+\rho^s)\in(M+h^s)Q$ for any $w,w'\in W$. This follows line for line the proof of Theorem 7.1 of \cite{MPCubature} since it does not change anything if we consider $h^s$ instead of $h$.

For the last line of the statement use the definition \eqref{kappa} of $\kappa^{s}$.

We can prove the result for the long root case similarly. However, there is one difference
which arises because $h^{s}= 1+\sum_{\alpha_i\in\Pi^s}m_i^\vee$ 
whereas $ h^l=\sum_{\alpha_i\in\Pi^l}m_i^\vee$.
This difference appears in the validation of the separation conditions, which hold only
for $\sum m_i^\vee \lambda_i\leq M$ and $\sum m_i^\vee \mu_i\leq M-1$ in the long case. 
\end{proof}

\subsection{The cubature formulas}\

The following Theorem can be proved in the same way as Theorem \ref{thmformule} since $\mu=0$ and $\lambda$ with $\sum \lambda_im_i^\vee\leq 2M+1$ ($2M-1$ respectively) satisfy the separation conditions of Prop.~\ref{s-discOrthogonality}. 
\begin{thm}\label{slcubature}\
\begin{enumerate}
\item[i)] Let $M\in\Z^{\geq0}$ and $f$ be any polynomial in $\C[X_1^s,\dots,X_n^s]$ with $m-deg(f)\leq 2M+1$, then 
\begin{equation*}
\begin{aligned}
\int_{\Omega^s}f (K^s)^{1/2}dX_1^s\dots dX_n^s=\frac{1}{c_\mathfrak{g}}\left(\frac{2\pi}{M+h^s}\right)^n\sum_{X^{s}\in\mathcal{F}^s_{M+h^s}}f(X^{s})\kappa^{s}(X^{s})\,,
\end{aligned}
\end{equation*}
where $\kappa^s$ is defined by \eqref{kappa}. 
\item[ii)] Let $M\in\N$ and $f$ be any polynomial in $\C[X_1^l,\dots,X_n^l]$ with $m-deg(f)\leq 2M-1$, then 
\begin{equation*}
\begin{aligned}
\int_{\Omega^l}f (K^l)^{1/2}dX_1^l\dots dX_n^l=\frac{1}{c_\mathfrak{g}}\left(\frac{2\pi}{M+h^l}\right)^n\sum_{X^{l}\in\mathcal{F}^l_{M+h^l}}f(X^{l})\kappa^{l}(X^{l})\,.
\end{aligned}
\end{equation*}
\end{enumerate} 
\end{thm}

\begin{remark}
One notes here that the short root case (i) is Gaussian cubature, with maximal efficiency in terms of the number of nodal points required, while the long root case (ii) fits into the  Radau cubature class and is slightly less efficient.
\end{remark}

\section{Approximating functions on $\Omega^s$ and $\Omega^l$}\

In this section we just point out a few things that are direct consequences of the 
Fourier analysis that has been developed here. As usual, we write this down for the short
root length case, the long root case being entirely parallel.

\subsection{Polynomial expansion in terms of $\chi^s_\lambda$}\

Let $L^2_{K^s}(\Omega^s)$ denote the space of all complex valued functions $f$ on $\Omega^s$ such that $\int_{\Omega^s}|f|^2 \,(K^{s})^{1/2} d X^{s}<\infty$. We recall the inner product 
of \eqref{transition} on $L^2_{K^s}(\Omega^s)$ 
\begin{equation*}
(f,g)_s:=(2\pi)^{-n}\int_{\Omega^s}f(X^{s})\overline{g(X^{s})}(K^{s}(X^{s}))^{1/2}dX^{s}
=\int_{F^{s}} f(X^s(x)) \overline{g(X^s(x))} S_{\rho^{s}}^{s}(x)\overline{S_{\rho^{s}}^{s}(x)}d\theta_{\T}(x)\,.
\end{equation*}

We write $f\bumpeq g$ if $f=g$ almost everywhere in $\Omega^s$. Since $(K^{s})^{1/2}$ is continuous and strictly positive on interior of $\Omega^s$, we have for any $f$ that $( f,f)_s\geq0$ with equality if and only if $f\bumpeq 0$. Thus, we can regard $L^2_{K^s}(\Omega^s)$ as a Hilbert space with $L^2_{K^s}-$norm of $f$ equal to $( f,f)_s^{1/2}$.

By Proposition \ref{s-cntsOrthogonality}, the polynomials 
$X^s_{\lambda}:=\chi^s_{\lambda}(x),\,x\in F^s$ with $\lambda\in P^+$ form an orthogonal set in $L^2_{K^s}(\Omega^s)$:
$$( X_{\lambda}^s,X_\mu^s)_s=|\mbox{stab}_W(\lambda+\rho^s)|^{-1}\delta_{\lambda\mu}\, ,$$
and, in fact, they form a Hilbert basis in $L^2_{K^s}(\Omega^s)$. We can see this by relating $f(X^{s})$ on $\Omega^s$ with $f(x)$ on $F^s$ and using the discussion in Section \ref{inp} to make its
Fourier expansion. Rewriting this back in $\Omega^{s}$ we obtain the basic expansion formulas
\begin{equation*}
f\bumpeq\sum_{\lambda\in P^+}a_\lambda X^s_\lambda\,,\qquad \text{where }a_\lambda=|\mbox{stab}_W(\lambda+\rho^s)|( f,X^s_\lambda)_s\,.
\end{equation*}

\subsection{Optimality}\

If $|\lambda|_m :=\sum_i m_i^\vee \lambda_i$, then the sums
$$\sum_{|\lambda|_m\leq M}|\mbox{stab}_W(\lambda+\rho^s)|( f,X^s_\lambda)_s X^s_\lambda$$
are the polynomials of $m-$degree at most $M$ in the variables $X_1^s,\dots, X^s_n$.

\begin{tvr}
Let $f\in L^2_{K^s}(\Omega^s)$. Amongst all polynomials $p(X_1^s,\dots,X_n^s)$ of $m-$degree less than or equal
to $M$, the polynomial
$q=\sum_{|\lambda|_m\leq M}|\mbox{stab}_W(\lambda+\rho^s)|( f,X^s_\lambda)_s X^s_\lambda$ is the best approximation to $f$ relative to the $L^2_{K^s}-$norm.
\end{tvr}

\begin{proof}
Let $p=\sum_{|\lambda|_m\leq M}b_\lambda X^s_\lambda$ be any polynomial of $m-$degree at most $M$ and $a_\lambda=|\mbox{stab}_W(\lambda+\rho^s)|( f,X^s_\lambda)_s$, then
\begin{equation*}
\begin{aligned}
( f-p,f-p)_s&=( f,f)_s-\sum_{|\lambda|_m\leq M}|\mbox{stab}_W(\lambda+\rho^s)|^{-1} a_\lambda\overline{b_\lambda}-\sum_{|\lambda|_m\leq M}|\mbox{stab}_W(\lambda+\rho^s)|^{-1}b_\lambda \overline{a_\lambda}\\&+\sum_{|\lambda|_m\leq M}|\mbox{stab}_W(\lambda+\rho^s)|^{-1}|b_\lambda|^2=( f-q,f-q )_s+\sum_{|\lambda|_m\leq M}|\mbox{stab}_W(\lambda+\rho^s)|^{-1}|b_\lambda-a_\lambda|^2\\&\geq ( f-q,f-q )_s
\end{aligned}
\end{equation*}
with equality if and only if $b_\lambda=a_\lambda$.
\end{proof}

\section{Example: Cubature formulas for $G_2$}\
In this section we illustrate briefly how the main constituents of the paper look in the
case of the Lie group $G_{2}$ when $M=15$. 

\subsection{$S^s$- and $S^l$-functions of $G_2$}\

Let us recall some basic facts about Lie group $G_2$.  The simple roots  $\alpha_1,\alpha_2$ and co-roots $\alpha_1^\vee,\alpha_2^\vee$ are determined by the Cartan matrices $C$ and $C^T$;
\begin{equation*}
C=\left(\begin{matrix}
2&-3\\-1&2
\end{matrix}\right)\,,\quad C^T=\left(\begin{matrix}
2&-1\\-3&2
\end{matrix}\right)\,.
\end{equation*}
We also have the following relations between the bases:
\begin{equation*}
\begin{alignedat}{5}
&\alpha_1=2\omega_1-3\omega_2\,,&\quad &\alpha_2=-\omega_1+2\omega_2\,,&\quad &\omega_1=2\alpha_1+3\alpha_2\,,&\quad&\omega_2=\alpha_1+2\alpha_2\,;\\
&\alpha_1^\vee=2\omega^\vee_1-\omega^\vee_2\,,&\quad &\alpha^\vee_2=-3\omega^\vee_1+2\omega^\vee_2\,,&\quad &\omega^\vee_1=2\alpha^\vee_1+\alpha^\vee_2\,,&\quad&\omega^\vee_2=3\alpha^\vee_1+2\alpha^\vee_2\,.
\end{alignedat}
\end{equation*}
Using \eqref{h-values}, $\rho^{s} = \omega_{2}$, $\rho^{l} = \omega_{1}$, $h^{s }= h^{l}= 3$.

\begin{figure}
\centering
\includegraphics[width=8cm]{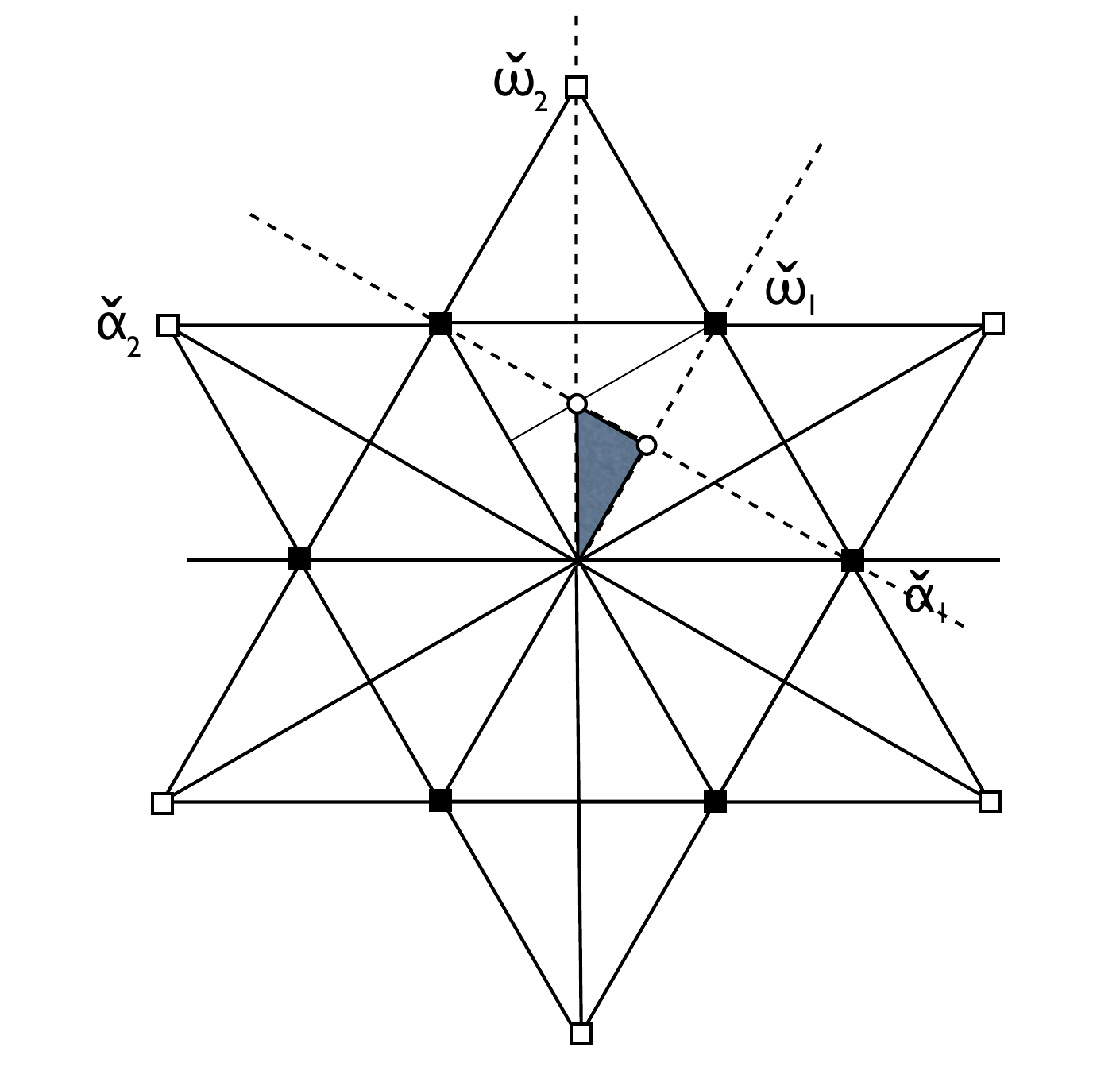}
\caption{A schematic view of the co-root system of $G_2$. The shaded triangle is the
fundamental region $F$. The dotted lines are the mirrors which define its boundaries, the
reflections in which generate the affine Weyl group. The action of the affine Weyl group 
on $F$ tiles the plane.  A few tiles of this tiling are shown. Filled (resp, open) squares are the short (resp. long) co-roots of $G_2$.}\label{G2figure}
\label{g2}
\end{figure} 

The defining relations for the Weyl group are $r_{1}^{2}=r_{2}^{2}= (r_{1}r_{2})^{6}=1$.
Defining $r_{opp}:= r_{1}r_{2}r_{1}r_{2}r_{1}r_{2}= r_{2}r_{1}r_{2}r_{1}r_{2}r_{1}$, 
the Weyl group consists of
$1, r_{1},r_{2},r_{1}r_{2}, r_{2}r_{1}, r_{1}r_{2}r_{1}$, together with the product of $r_{opp}$ with each of these elements. The corresponding values of $\sigma^{s}$ are 
$1,1,-1,-1,-1,-1$ and $\sigma^{s}(r_{opp}) = -1$; and for 
$\sigma^{l}$ they are $1,-1,1,-1,-1,1$ and $\sigma^{l}(r_{opp}) = -1$.

Let $\lambda=(\lambda_1,\lambda_2)=\lambda_1\omega_1+\lambda_2\omega_2$ and $x=(x_1,x_2)=x_1\alpha_1^\vee+x_2\alpha_2^\vee$. Any Weyl group orbit of a generic point $\lambda$ consists of
$$
\{\pm(\lambda_1,\lambda_2)\,,\pm(-\lambda_1,3\lambda_1+\lambda_2)\,,\pm(\lambda_1+\lambda_2,-\lambda_2)\,,\pm(2\lambda_1+\lambda_2,-3\lambda_1-\lambda_2)\,,$$ $$\pm(-\lambda_1-\lambda_2,3\lambda_1+2\lambda_2)\,,\pm(-2\lambda_1-\lambda_2,3\lambda_1+2\lambda_2)\}\,.$$
Therefore the explicit formulas for the $S^s$- and $S^l$-functions are:
\begin{alignat*}{2}
S^s_{\lambda+\omega_2}(x)&=\frac{2i}{|\mathrm{stab}_W(\lambda+\omega_2)|}(\sin{2\pi(\lambda_1 x_1+(\lambda_2+1) x_2)}
           +\sin{2\pi(-\lambda_1x_1+(3\lambda_1+\lambda_2+1)x_2)}\\
            &-\sin{2\pi((\lambda_1+\lambda_2+1)x_1-(\lambda_2+1)x_2)}
           -\sin{2\pi((2\lambda_1+\lambda_2+1)x_1+
           (-3\lambda_1-\lambda_2-1)x_2)}\\
            &-\sin{2\pi((-\lambda_1-\lambda_2-1)x_1+(3\lambda_1+2\lambda_2+2)x_2)}\\&  
           -\sin{2\pi((-2\lambda_1-\lambda_2-1)x_1+(3\lambda_1+2\lambda_2+2)x_2)}\,,\\
S^l_{\lambda+\omega_1}(x)&=\frac{2i}{|\mathrm{stab}_W(\lambda+\omega_1)|}(\sin{2\pi((\lambda_1+1)x_1+\lambda_2x_2)}
           -\sin{2\pi(-(\lambda_1+1)x_1+(3\lambda_1+\lambda_2+3)x_2)}\\
            &+\sin{2\pi((\lambda_1+\lambda_2+1)x_1-\lambda_2x_2)}
           -\sin{2\pi((2\lambda_1+\lambda_2+2)x_1+(-3\lambda_1-\lambda_2-3)x_2)}\\
            &-\sin{2\pi((-\lambda_1-\lambda_2-1)x_1+(3\lambda_1+2\lambda_2+3)x_2)} 
            \\& 
           +\sin{2\pi((-2\lambda_1-\lambda_2-2)x_1+(3\lambda_1+2\lambda_2+3)x_2)}\,.
\end{alignat*}
By definition the polynomial variables $X_1^s,X_2^s$ and $X_1^l,X_2^l$ are given by 
\begin{equation}\label{polG2}
\begin{alignedat}{2}
X_1^s=\frac{S^s_{(1,1)}(x)}{S^s_{(0,1)}(x)}&=2 (1 + \cos{2\pi x_1} + \cos{2 \pi (x_1 - 3 x_2)} + 
   2 \cos{2 \pi (x_1 - 2 x_2)} + 2 \cos{2 \pi (x_1 - x_2)}\\& + 
   2 \cos{2 \pi x_2}  
   +\cos{2 \pi (2x_1 - 3 x_2)})\,,\\
X_2^s=\frac{S^s_{(0,2)}(x)}{S^s_{(0,1)}(x)}&=2(1 + \cos{2 \pi x_2} + \cos{2 \pi (x_1 - 2 x_2)} + \cos{2 \pi (x_1 - x_2)})\,;\\
X_1^l=\frac{S^l_{(2,0)}(x)}{S^l_{(1,0)}(x)}&=2 (1 + \cos{2 \pi x_1} + \cos{2 \pi (x_1 - 3 x_2)} + \cos{2 \pi (2 x_1 - 3 x_2)})\,,\\
X_2^l=\frac{S^l_{(1,1)}(x)}{S^l_{(1,0)}(x)}&=2 (\cos{2 \pi x_2} + \cos{2 \pi (x_1 - 2 x_2)} + \cos{2 \pi (x_1 - x_2)})\,.   
\end{alignedat}
\end{equation}

\subsection{Integration regions $\Omega^s,\Omega^l$ and grids $\mathcal{F}_{M+3}^s,\mathcal{F}_{M+3}^l$}\

\begin{figure}[h]
\includegraphics[width=0.8\textwidth]{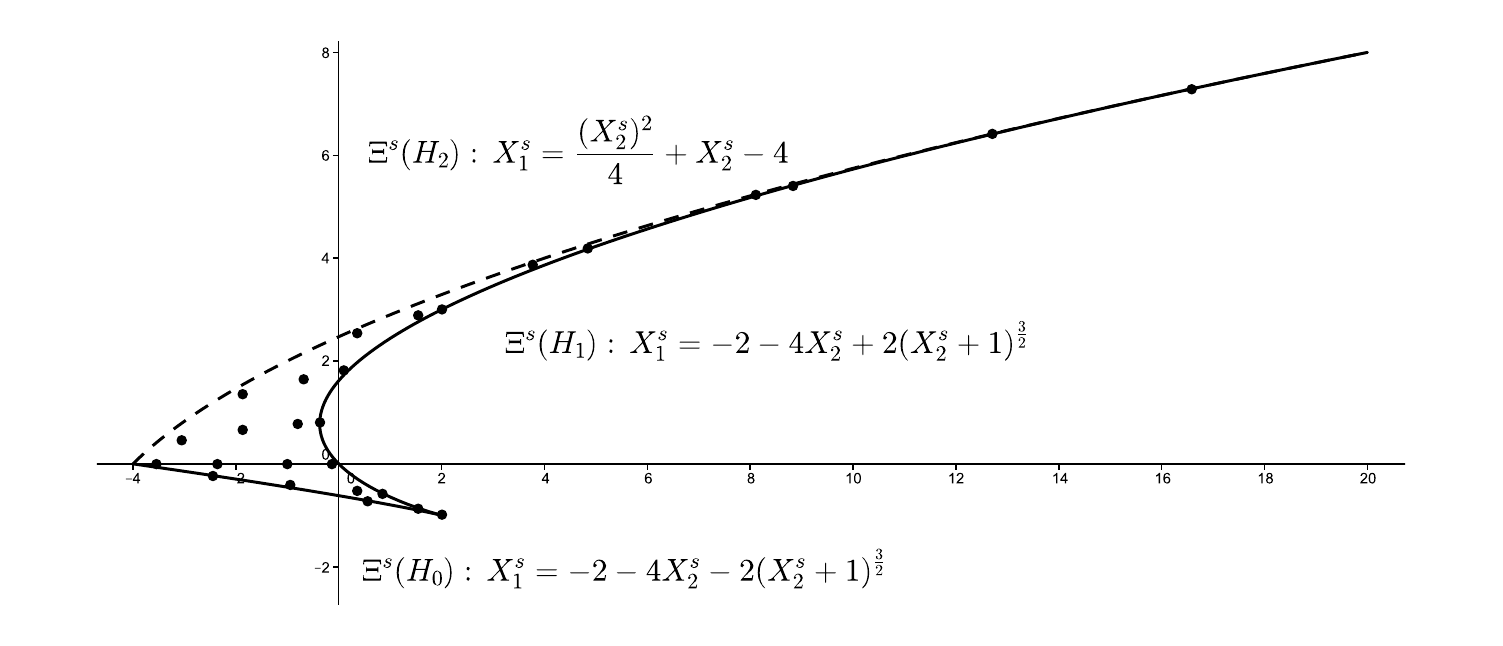}
\caption{The region $\Omega^s$ along with the equations of
its boundaries. Inside we see the
points of $\mathcal{F}_{18}^s$. The dashed boundary is not included in $\Omega^s$.}
\label{reg1} 
\end{figure}

\begin{figure}[h]
\includegraphics[width=0.45\textwidth]{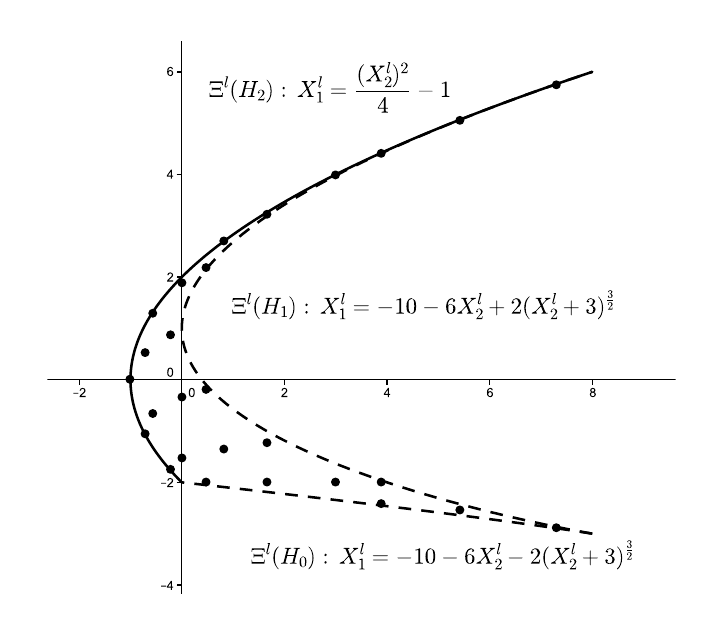}
\caption{The region $\Omega^l$ along with the equations of
its boundaries. Inside we see the
points of $\mathcal{F}_{18}^l$. The dashed boundaries are not included in $\Omega^l$.}
\label{reg2} 
\end{figure}

Using the explicit formulas \eqref{polG2} for polynomial variables as functions of $x_1,x_2$, one can determine the integration regions $\Omega^s,\Omega^l$ (see Figure \ref{reg1} and \ref{reg2}), namely:
\begin{align*}
\Omega^s&=\set{(X_1^s,X_2^s)}{X_1^s> \frac{(X_2^s)^2}{4}+X^s_2-4\,, -2-4X_2^s-2(X_2^s+1)^\frac{3}{2}\leq X_1^s\leq -2-4X_2^s+2(X_2^s+1)^\frac{3}{2}}\,;\\
\Omega^l&=\set{(X_1^l,X_2^l)}{X_1^l\geq \frac{(X_2^l)^2}{4}-1\,, -10-6X_2^l-2(X_2^l+3)^\frac{3}{2}<X_1^l< -10-6X_2^l+2(X_2^l+3)^\frac{3}{2}}\,.
\end{align*}

The grids $\mathcal{F}_{M+3}^s,\mathcal{F}_{M+3}^l$ are the following finite sets of points in $\Omega^s$ and $\Omega^l$ respectively.
 \begin{align*}
\mathcal{F}_{M+3}^s&=\left\{\left(X_1^s\left(\frac{2s_1+3s_2}{M+3},\frac{s_1+2s_2}{M+3}\right),X_2^s\left(\frac{2s_1+3s_2}{M+3},\frac{s_1+2s_2}{M+3}\right)\right)\right\}\,,\\
&\text{where }s_1=0,\dots,\left\lfloor\frac{M+3}{2}\right\rfloor,s_2=1,\dots,\left\lfloor\frac{M+3-2s_1}{3}\right\rfloor\,;\\
\mathcal{F}_{M+3}^l&=\left\{\left(X_1^l\left(\frac{2s_1+3s_2}{M+3},\frac{s_1+2s_2}{M+3}\right),X_2^l\left(\frac{2s_1+3s_2}{M+3},\frac{s_1+2s_2}{M+3}\right)\right)\right\}\,,\\
&\text{where }s_1=1,\dots,\left\lfloor\frac{M+2}{2}\right\rfloor,s_2=0,\dots,\left\lfloor\frac{M+2-2s_1}{3}\right\rfloor\,.
\end{align*}

The list of EFOs for $M=15$ is given in Table~\ref{EFO18List}.
\begin{table}
\begin{center} 
\small
\begin{tabular}{c|c|c|c|c}
$(s_0,s_1,s_2)$&$F_{18}^s$&$F_{18}^l$&$(X_1^s,X_2^s)$&$(X_1^l,X_2^l)$\\\hline
$(0, 0, 6)$&\checkmark&$\times$&$(2,-1)$&$\times$\\ 
$(0, 3, 4)$&\checkmark&$\times$&$(0.5662, -0.7169)$&$\times$\\
$(0, 6, 2)$&\checkmark&$\times$&$(-2.4534, -0.2267)$&$\times$\\
 $(0, 9, 0)$&$\times$&$\times$&$\times$&$\times$\\
 $(1, 1, 5)$&\checkmark&\checkmark&$(1.5321, -0.8794)$&$(7.2909, -2.8794)$\\
 $(1, 4, 3)$&\checkmark&\checkmark&$(-0.9436, -0.4115)$&$(3.8794, -2.4115)$\\
 $(1,7, 1)$&\checkmark&\checkmark&$(-3.5321, 0)$&$(0.4679, -2)$\\ 
 $(2, 2, 4)$&\checkmark&\checkmark&$( 0.3473, -0.5321)$&$(5.4115, -2.5321)$\\
 $(2, 5, 2)$&\checkmark&\checkmark&$(-2.3473, 0)$&$( 1.6527, -2)$\\
 $(2, 8, 0)$&$\times$&\checkmark&$\times$&$(-0.2267, -1.7588)$\\
 $(3, 0, 5)$&\checkmark&$\times$&$(0.852, -0.574)$&$\times$\\
 $(3, 3, 3)$&\checkmark&\checkmark&$(-1,0)$&$(3,-2)$ \\
 $(3, 6,1)$ &\checkmark&\checkmark&$(-3.0642,0.4679)$&$(0, -1.5321)$\\
 $(4, 1, 4)$&\checkmark&\checkmark&$(-0.1206,0)$&$(3.8794, -2)$\\
 $(4, 4, 2)$&\checkmark&\checkmark&$(-1.8794, 0.6527)$&$( 0.8152, -1.3473)$\\
 $(4, 7, 0)$&$\times$&\checkmark&$\times$&$( -0.7169, -1.0642)$\\
 $(5, 2, 3)$&\checkmark&\checkmark&$(-0.8007, 0.7733)$&$( 1.6527, -1.2267)$\\
 $(5, 5, 1)$&\checkmark&\checkmark&$(-1.8794, 1.3473)$&$(-0.574, -0.6527)$\\
 $(6, 0, 4)$&\checkmark&$\times$&$(-0.3696,0.8152)$&$\times$\\
 $(6, 3, 2)$&\checkmark&\checkmark&$(-0.6946, 1.6527)$&$(0, -0.3473)$\\
 $(6, 6, 0)$&$\times$&\checkmark&$\times$&$(-1,0)$\\
 $(7, 1, 3)$&\checkmark&\checkmark&$(0.0983, 1.8152)$&$(0.4679, -0.1848)$\\
 $(7, 4, 1)$&\checkmark&\checkmark&$(0.3473,2.5321)$&$(-0.7169, 
  0.5321)$\\
 $(8, 2, 2)$&\checkmark&\checkmark&$(1.5321, 2.8794)$&$(-0.2267, 0.8794)$\\
 $(8, 5, 0)$&$\times$&\checkmark&$\times$&$(-0.574, 
  1.3054 )$\\
 $(9, 0, 3)$&\checkmark&$\times$&$(2,3)$&$\times$\\
 $(9, 3, 1)$&\checkmark&\checkmark&$(3.7588, 3.8794)$&$(0, 1.8794)$\\
 $(10, 1, 2)$&\checkmark&\checkmark&$(4.8375,4.1848)$&$(0.4679, 2.1848)$\\
 $(10, 4, 0)$&$\times$&\checkmark&$\times$&$( 0.8152, 2.6946)$\\
 $(11, 2, 1)$&\checkmark&\checkmark&$(8.1061, 5.2267)$&$( 1.6527, 3.2267)$\\
 $(12, 0, 2)$&\checkmark&$\times$&$(8.823, 5.4115)$&$\times$\\
 $(12, 3, 0)$&$\times$&\checkmark&$\times$&$(3,4)$\\
  $(13, 1, 1)$&\checkmark&\checkmark&$(12.7023, 6.4115)$&$(3.8794, 
  4.4115)$\\
 $(14, 2, 0)$&$\times$&\checkmark&$\times$&$(5.4115, 5.0642)$\\
 $(15, 0, 1)$&\checkmark&$\times$&$(16.5817, 7.2909)$&$\times$\\
 $(16, 1, 0)$&$\times$&\checkmark&$\times$&$(7.2909, 
  5.7588)$\\
 $(18, 0, 0)$&$\times$&$\times$&$\times$&$\times$
\end{tabular}
\medskip
\caption{A list of the EFOs for $M+3 = 18$, along with their coordinates
in the domains $\Omega^{s}$ and $\Omega^{l}$. Since $F^{s}$ is missing the boundary
defined by the fixed hyperplane for the short reflection $r_{2}$, EFOs falling on this boundary
are not part of the short root scenario. For $F^{l}$ it is EFOs on the hyperplanes for $r_{1}$ and $r_{0}$
that are not included.}
\label{EFO18List}
\end{center}
\end{table}
\subsection{Cubature formulas}\

The functions $K^s$ and $K^l$ are given by the expressions:
\begin{align*}
K^s(X_1^s,X_2^s&)=\frac{S^s_{\omega_2}\overline{S^s_{\omega_2}}}{S^l_{\omega_1}\overline{S^l_{\omega_1}}}=\frac{-(X_2^s)^2-4X_2^s+4X_1^s+16}{4(X_2^s)^3-(X_1^s)^2-4(X_2^s)^2-8X_1^sX_2^s-4X_1^s-4X_2^s}\,;\\
K^l(X_1^l,X_2^l&)=\frac{S^l_{\omega_1}\overline{S^l_{\omega_1}}}{S^s_{\omega_2}\overline{S^s_{\omega_2}}}=\frac{4(X_2^l)^3-(X_1^l)^2-12X_1^lX_2^l-20X_1^l-12X_2^l+8}{-(X_2^l)^2+4X_1^l+4}\,.
\end{align*}
Thus, the explicit cubature formulas of $G_2$ are
\begin{align*}
\int_{\Omega^s}&f(X_1^s,X_2^s)\sqrt{\frac{-(X_2^s)^2-4X_2^s+4X_1^s+16}{4(X_2^s)^3-(X_1^s)^2-4(X_2^s)^2-8X_1^sX_2^s-4X_1^s-4X_2^s}}\,dX^s_1\,dX^s_2\\&=
\frac{1}{12}\left(\frac{2\pi}{M+3}\right)^2\sum_{(X_1^s,X_2^s)\in\mathcal{F}_{M+3}^s}f(X_1^s,X_2^s)|Wx|(-(X_2^s)^2-4X_2^s+4X_1^s+16)\,;\\
\int_{\Omega^l}&f(X_1^l,X_2^l)\sqrt{\frac{4(X_2^l)^3-(X_1^l)^2-12X_1^lX_2^l-20X_1^l-12X_2^l+8}{-(X_2^l)^2+4X_1^l+4}}\,dX^l_1\,dX^l_2\\&=
\frac{1}{12}\left(\frac{2\pi}{M+3}\right)^2\sum_{(X_1^l,X_2^l)\in\mathcal{F}_{M+3}^l}f(X_1^l,X_2^l)|Wx|(4(X_2^l)^3-(X_1^l)^2-12X_1^lX_2^l-20X_1^l-12X_2^l+8)\,.
\end{align*}
The values $|Wx|$ are written in Table \ref{eps}.
\begin{table}
\begin{tabular}{c|c}
$\left(s_0,s_1,s_2\right)$&$|Wx|$\\
\hline
$(\star,0,0)$&1\\
$(0,0,\star)$&2\\
$(0,\star,0)$&3\\
$(0,\star,\star)$&6\\
$(\star,0,\star)$&6\\
$(\star,\star,0)$&6\\
$(\star,\star,\star)$&12
\end{tabular}
\medskip
\caption{A table of values of $|Wx|$ for the group $G_2$ based on the form of the 
coordinates of $x \in F$. Recall that this is a count of the $W$ orbit of $x$ taken 
modulo $Q^{\vee}$. The values can be worked out using Fig.~\ref{g2}. The cases
$(\star,0,0),(0,\star, 0)$ do not appear in this context, but we include them to complete
the table.}
\label{eps}
\end{table}
\section*{Acknowledgements}\

We gratefully acknowledge the support of this work by the Natural Sciences and Engineering Research Council of Canada and by the MIND Research Institute of Irvine, Calif. L.M. would also like to express her gratitude to the Centre de recherches math\'ematiques, Universit\'e de Montr\'eal, for the hospitality extended to her during her doctoral studies as well as to the Institute de Sciences 
Math\'ematiques de Montr\'eal and Foundation J.A. DeS\`eve for partial support of her studies.


\begin{thebibliography}{999}

\small
\bibitem{MPCubature}
Moody, R.V., Patera, J.: {\it Cubature formulae for orthogonal polynomials in terms of
elements of finite order of compact simple Lie groups}. Advances in Applied Mathematics 47, 509-535 (2011)

\bibitem{LX}
Li, H., Xu, Y.: {\it Discrete Fourier analysis on fundamental domain and simplex of Ad lattice in d-variables}, J. Fourier Anal. Appl. 16,  383-433, (2010)

\bibitem{hec}
Heckman, G., Schlichtkrull, H.: {\it Harmonic Analysis and Special Functions on Symmetric Spaces.\/} Academic Press Inc., San Diego (1994)

\bibitem{B}
Bourbaki, N.: {\it Groupes et alg\`ebres de Lie, Ch. 4,5,6,\/} {\sl \'El\'ements de Math\'ematiques}. Hermann, Paris (1968)

\bibitem{mpsk}
Moody, R.V., Patera, J., Slansky, R.: {\it Affine Lie Algebras, Weight Multiplicities, and Branching Rules.} Vol. 1, University of California Press (1990)

\bibitem{PSS}  
Patera, J., Sharp, R.T., Slansky, R.: {\it On a new relation between semisimple Lie algebras.\/}  J.~Math.~Phys. {\bf 21}, 2335-2341 (1980)

\bibitem{MPi}  
Moody, R.V., Pianzola, A.: {\it $\lambda$-mapping between representation rings of Lie algebras.\/} Canad. J. Math. {\bf 35}, 898-960 (1983)

\bibitem{Serre}
Serre, J.-P.: \textit{Alg\`ebres de Lie Semi-simples Complexes}. Benjamin, (1966)
[English trans. \textit{Complex Semisimple Lie Algebras}. Springer (2001)]


\bibitem{KP06}
Klimyk, A.U., Patera, J.: {\it Orbit functions.\/}  SIGMA (Symmetry,
Integrability and Geometry: Methods and Applications) {\bf 2}, Paper 006, 60pp. (2006)

\bibitem{KP07}
 Klimyk, A.U., Patera, J.: {\it Antisymmetric orbit functions.\/}  SIGMA {\bf 3}, Paper 023, 83 pp. (2007)


\bibitem{MP84}  
Moody, R.V., Patera, J.: {\it Characters of elements of finite order in simple Lie groups.\/} SIAM J. on Algebraic and Discrete Methods {\bf 5}, 359-383 (1984)


\bibitem{HMP}
Hrivn{\'a}k, J., Motlochov{\'a}, L., Patera, J.: {\it On Discretization of Tori of Compact Simple Lie Groups II..\/} J. Phys. A {\bf 45}, 255201 (2012)




\end{thebibliography}
\end{document}